\documentclass[aps,preprint]{revtex4}
\usepackage[T1]{fontenc}
\usepackage[utf8]{inputenc}
\usepackage{array}
\usepackage{multirow}
\usepackage{amsmath}
\usepackage{amssymb}
\usepackage{graphicx}
\usepackage{makecell}
\usepackage{color}
\usepackage{comment}
\usepackage{amsmath} 
\usepackage{booktabs} 

\makeatletter

\@ifundefined{textcolor}{}
{%
	\definecolor{BLACK}{gray}{0}
	\definecolor{WHITE}{gray}{1}
	\definecolor{RED}{rgb}{1,0,0}
	\definecolor{GREEN}{rgb}{0,1,0}
	\definecolor{BLUE}{rgb}{0,0,1}
	\definecolor{CYAN}{cmyk}{1,0,0,0}
	\definecolor{MAGENTA}{cmyk}{0,1,0,0}
	\definecolor{YELLOW}{cmyk}{0,0,1,0}
}

\usepackage{amsfonts}
\usepackage{array}
\usepackage{multirow}
\usepackage{latexsym}
\usepackage{epsfig}
\usepackage{float}
\usepackage{dsfont}

\usepackage{ragged2e}
\justifying\let\raggedright\justifying
\usepackage{caption}
\newcommand{\be}{\begin{equation}}
	\newcommand{\ee}{\end{equation}}

\RequirePackage{doi}

\makeatother

\begin{document}
	\title{Superoscillations in High Energy Physics and Gravity}
	
	\author{Andrea Addazi}
	\email{addazi@scu.edu.cn}
	\affiliation{Center for Theoretical Physics, College of Physics Science and Technology, Sichuan University, 610065 Chengdu, China}
	\affiliation{Laboratori Nazionali di Frascati INFN, Frascati (Rome), Italy, EU}
	
	\author{Qingyu Gan}
	\affiliation{Scuola Superiore Meridionale, Largo S. Marcellino 10, I-80138, Napoli,
		Italy}
	
	\vspace{1cm}

	\begin{abstract}

		We explore superoscillations within the context of classical and quantum field theories, presenting novel solutions to Klein-Gordon's, Dirac's, Maxwell's and Einstein's equations. 
		In particular, we illustrate a procedure of second quantization of fields
		and how to construct a Fock space which encompasses Superoscillating states. 
		Furthermore, we extend the application of superoscillations to quantum tunnelings, scatterings and mixings of particles, squeezed states and potential advancements in laser interferometry, which could open new avenues for experimental tests of Quantum Gravity effects. 
		By delving into the relationship among superoscillations and phenomena such as Hawking radiation, the Black Hole (BH) information and the Firewall paradox, we propose an alternative mechanism for information transfer  across the BH event horizon.

	\end{abstract}
	\maketitle
	\tableofcontents{}
	
	\vspace{3cm}
	
	\section{Introduction}
	
	Superoscillation is a remarkable phenomenon wherein a band-limited function exhibits oscillations faster than its fastest Fourier component. Originally predicted by {\it Aharanov et al} and {\it Berry}, this behavior has been experimentally observed in laboratories using optical  microscopes (see Ref.\cite{Review} for a comprensive review).
	
	From a theoretical standpoint, superoscillations have been extensively studied in non-relativistic Quantum Mechanics (QM), where they arise as solutions to Schr\"odinger’s equation \cite{Super1,Super2,Super3,Super4,Osc1,Osc2,Review}.
	
	\vspace{0.2cm}
	
	In this paper, we propose a novel formulation of superoscillation within the context of Field Theories. We will present superoscillating solutions for spin 0, 1/2, 1, and 2 fields, beginning with an exploration of scalar field theory as a foundational step before delving into electrodynamics and General Relativity.
	Moreover, we will explore potential applications of superoscillations in various domains, including particle mixings and scatterings, laser interferometers for tests of Gravitational Waves and Quantum Gravity phenomena, Black Hole radiation. 
	
	In the subsequent section, we will provide a concise review of fundamental aspects of superoscillation in non-relativistic Quantum Mechanics and  we will add a novel discussion on Feynman's reformulations of quantum superoscillations from path integral prospective. Then, we will transition to a discussion on superoscillations in the context of field theories and their diverse applications.

	\vspace{0.1cm}
	
	\subsection{Superoscillations in  non-relativistic quantum mechanics }

	The most studied (but not unique) super-oscillating function  \cite{Super1,Super2,Super3,Super4,Osc1,Osc2,Review} is
	$$F_{n}(x;a)=\Big(\cos \frac{x}{n}+ia\sin\frac{x}{n}\Big)^{n},\,\,\, a>1$$
	$$=\sum_{k=0}^{n}c_{k}(n;a)e^{ix\left(1-\frac{2 k}{n}\right)},\,\,\,\ $$
	\begin{equation}
		\label{super}
		c_{k}(n;a)=\frac{n!}{k!(n-k)!}\Big( \frac{1+a}{2}\Big)^{n-k}\Big( \frac{1-a}{2}\Big)^{k}\, ,
	\end{equation}
	where $x$ is normalised as an adimensional variable
	and $n$ is a large even integer \footnote{There are others well-known superoscillating functions, the applications of which could potentially be expanded to Quantum Field Theories. For example, 
		a monocromatic $\psi_{m}(r,\phi)=J_{m}(r)e^{im\phi}+\epsilon J_{0}(r)$ studied in context of phase singularities related to nodal points or lines or wave dislocations etc \cite{BesselJJJ}. 
		Another interesting superoscillating profile is $F_{s}(x)=[\cos(\pi f x)-s]^{2}$
		with $0<s<1$, with highest Fourier frequency $f$, and around $x=0$ it contains as oscillation with width $t=(2/\pi f)\cos^{-1}s$ \cite{coss}.  
		Moreover, an antenna array approach for superoscillations can be also considered from N-isotropic sources with a radiation reading as 
		${\bf A}(\theta)=\sum_{n=0}^{N-1}c_{n}e^{i n k d\sin\theta}$,
		with $\theta$ the radiation angle, 
		$d$ the inter-separation of the antenna
		elements and $k_{0}$ the wave-number in free space \cite{Antenna}. 
	}. 
	
	Such a function has a period $n\pi$.
	It is a band-limited function and, if expanded in a Fourier series,
	the oscillation are all of the form $e^{ip_{k}x}$
	with $|p_{k}|\leq 1$ and $p_{k}=(1-2k/n)$.
	As it is known, Eq. \ref{super} can be approximated by 
	$e^{iax}$ within the interval $|x|<\sqrt{n}$.
	Outside this x-range, the function increases
	as anti-Gaussian and it rises to a maximum value 
	$F_{n}(\pm n\pi/2)=a^{n}$ to be ultimately destroyed outside the $|x|<n$ range. Indeed,  the $a$-parameter represents the superoscillation 
	degree while $n$ is the measure of the superoscillation region. 
	The speed of convergence is defined
	as follows:
	\begin{equation}
		\label{con}
		(F_{n}(x;a)-e^{iax})\sim \frac{x}{n}\sqrt{\frac{3}{2}(a^{2}-1)}\, .
	\end{equation}

\begin{figure}
		\centering
		\includegraphics[scale=0.88]{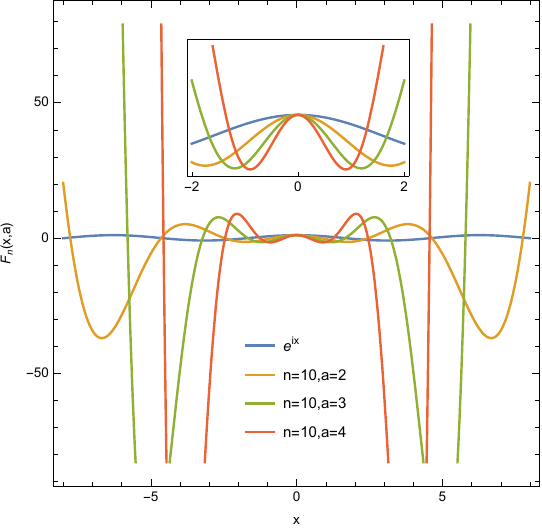}\includegraphics[scale=0.91]{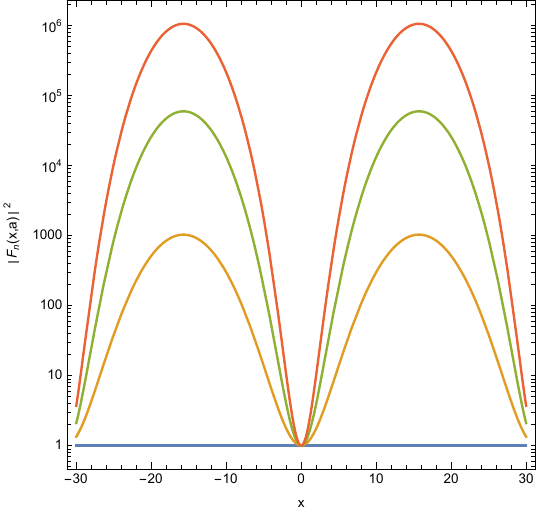}
		\caption{Several examples of superoscillation function Eq.\ref{super} with different parameters are displayed. }
		\label{fig:sup}
	\end{figure}

	As shown in Fig. \ref{fig:sup}, one can see that superoscillating functions oscillate faster than their maximal Fourier component in the limited region near $x=0$. The price to pay is  the suppression of the magnitude in this faster-oscillating region. 

	Considering Eq.\ref{super} as the initial wave function at $t=0$, 
	we can study its evolution in time. 
	The  case studied by {\it Aharonov et al} \cite{Super1,Super2,Super3,Super4,Osc1,Osc2,Review} corresponds to the $1+1$ Schr\"odinger equation: 
	\begin{equation}
		\label{Not}
		i\hbar \frac{\partial \psi}{\partial t}=\hat{H}\psi\, , 
	\end{equation}
	with 
	\begin{equation}
		\label{ff}
		\psi(x,0)=F_{n}(x;a)\, . 
	\end{equation}
	
	Starting from Eq.1 at $t=0$, 
	in the case of Schr\"odinger equation in vacuum with $\hat{H}=- \partial_x^2$ $(\hbar=1,m=1/2)$, one can construct the following solution: 
	\begin{equation}
		\label{pap}
		\psi_{n}(x,t)=\sum_{k=0}^{n}c_{k}(n,a)e^{i\left(1-\frac{2 k}{n}\right)x}e^{-i\left(1-\frac{2 k}{n}\right)^2t}\, . 
	\end{equation}
	In the limit of 
	$n\rightarrow \infty$, Eq.\ref{pap} 
	converges to 
	\begin{equation}
		\label{psi}
		\psi(x,t)\rightarrow e^{iax-ia^{2}t}\, . 
	\end{equation}
	
	Introducing physical dimensional quantity, the solution corresponds to 
	\begin{equation}
		\label{pap}
		\psi_{n}(x,t)=\sum_{k=0}^{n}c_{k}(n,a)e^{i\left(1-\frac{2 k}{n}\right){\bf p}\cdot {\bf x}/\hbar}e^{-i\left(1-\frac{2 k}{n}\right)^2 \, Et/ \hbar}
	\end{equation}
	where $E=|{\bf p}|^2/2m$. 
	and,
	for $n\rightarrow \infty$, one obtains   
	\begin{equation}
		\label{limit1}
		\rightarrow e^{ia{\bf p}\cdot{\bf x}/\hbar-ia^{2}Et/\hbar}\, . 
	\end{equation}
	The ${\bf p},E$ are the momentum and energy variables,
	while ${\bf x},t$ are dimensional space and time variables. 
	Moreover, such a solution can be generalised to $d+1$ space-time dimensions. 
	
	Eq.\ref{pap} can be interpret as a sum of n modes
	with energy $E_{k}=(1-2k/n)^2 E_{k}$ and momentum ${\bf p}_k=(1-2k/n){\bf p}$. 
	In the limit of $n\rightarrow \infty$, the whole superoscillation function 
	converges to a plane wave with $E_{a}=a^{2}E$ and ${\bf p}_{a}=a{\bf p}$.

	It is worth to remark that a superoscillating wave function,
	governed by Schr\"odinger's equation and band limited in $|{\bf p}\cdot {\bf x}|/\hbar<\sqrt{n}$, 
	will be evanescent for $Et/\hbar > \sqrt{n}$ \cite{Review}.

	Eq.\ref{pap} can be found 
	using Green's methods
	and the propagator as follows: 
	\begin{equation}
		\label{KK}
		\psi({\bf x},t)=\int d{\bf  x'} K({\bf x},t;{\bf x'},t')\psi({\bf x'},t')\, , 
	\end{equation}
	where the propagator $K$ is an integral Kernel between the initial and the final wave functions 
	$\psi({\bf x}',t')$ and $\psi({\bf x},t)$.
	
	Assuming that $t'=0$ and $\psi({\bf x}',0)|_{t=t'}=\psi_{n}({\bf x})$, setting the initial condition  $\psi({\bf x}',t')|_{t'=0}$ to be the form of Eq. \ref{pap}, 
	and that $\hat{H}=|\hat{{\bf p}}|^2/2m$, 
	we can obtain Eq.\ref{pap} 
	from 
	\begin{equation}
		\label{KK}
		\psi_{n}({\bf x},t)=\int d{\bf x'} K({\bf x},t;{\bf x'},t')\psi_{n}({\bf x'},t')|_{t'=0}\, . 
	\end{equation}
	It is worth to remark that the Propagator 
	and Green's function are the same for
	ordinary or superoscillation states.
	Indeed, the propagator of superoscillating waves satisfies the same equation of ordinary ones:
	\begin{equation}
		\label{prop}
		\Big(i\frac{\partial}{\partial t}-\hat{H}\Big)K(x-x')=\delta(x-x')\, . 
	\end{equation}

	In a path integral formulation, 
	Eq.\ref{KK} can also be rewritten 
	substituting 
	\begin{equation}
		\label{equationK}
		K({\bf x},t; {\bf x'},t')=\int_{{\bf x'}(t')}^{{\bf x}(t)}e^{i\mathcal{S}[{\bf y}]/\hbar}\mathcal{D}{\bf y}\, . 
	\end{equation}
	It is interesting to note that the mathematical definition of 
	the path integral is unchanged in case of superoscillating wave functions
	with respect to ordinary ones. 
	
	Propagator equations of standard quantum mechanics 
	are valid in superoscillation regime. 
	In particular, 
	the Fourier transform of the propagator corresponds to 
	$\tilde{K}({\bf p}',{\bf E}')=1/(E'-|{\bf p}'|^2/2m)$
	and the $\pm i \epsilon$
	conventions for retarded/advanced 
	Green functions are unchanged with respect to standard procedures for ordinary waves.  
	With superoscillation kinematic inputs 
	in the large-n limit, the energy and the momentum in the propagator rescale as $a^{2}$ and $a$ respectively.

	The inclusion of superoscillations in the context of Feynman's path integral does suggest a nuanced view of how paths contribute to the quantum amplitude and for instance in scenarios like the double-slit experiment. 
	Feynman's path integral formulation sums over all conceivable paths that a particle might take from one point to another, with each path contributing an amplitude weighted by the exponential of the action 
	$\mathcal{S}$ along that path, as expressed by $e^{i\mathcal{S}/\hbar}$. 
	This formulation inherently accounts for all possible trajectories, including those that might be considered part of superoscillatory behavior in the wave function.
	
	The impact of Superoscillations on path integral can be summarized as follows: 
	
	i) {\it Nature of Path Contributions}. Superoscillations do not modify the fundamental nature of the path integral itself; rather, they influence which paths contribute most significantly to the quantum amplitude. Superoscillating wave functions are characterized by rapid oscillations over small spatial regions, suggesting that paths corresponding to these high-frequency components can have significant contributions to the overall quantum behavior.
	
	ii) {\it Interference and Amplitude}. In the double-slit experiment, the interference pattern on the detection screen is the result of the constructive and destructive interference of amplitudes from paths through each slit. Superoscillations, being a wave phenomenon, imply that the paths associated with higher momenta (or more rapid oscillations) could contribute differently to the interference pattern, potentially leading to enhanced or altered interference effects compared to non-superoscillating wave functions.
	The presence of superoscillations affects the distribution and magnitude of quantum amplitudes across different paths due to the unique structure of the superoscillating wave function. This does not change the mechanics of how the path integral is computed but alters the contribution landscape of various paths due to the encoded high-frequency components.
	
	Thus, the path integral over infinite paths is not "modified" in the fundamental mathematical sense by the presence of superoscillations; the formulation remains the same, integrating over all paths with each weighted by $e^{i\mathcal{S}}$.  What changes is the effective contribution of different paths to the quantum amplitude, influenced by the initial conditions set by the superoscillating wave function. 
	
	Let us consider the simplified case of a double slit experiment.
	With both slits open, superoscillatory wave functions can produce an interference pattern on the detection screen that includes features not present in the interference pattern of ordinary wave functions.
	The superoscillatory nature of the wave function means that, in addition to the standard interference fringes, there may be regions with unexpectedly high-resolution fringes or localized intensity peaks, due to the high-frequency oscillations that are characteristic of superoscillations.
	These high-frequency components allow the wave function to "fit" through the slits in ways that might not be intuitive, potentially enhancing certain paths' contributions to the interference pattern due to constructive interference.
	Closing one slit while using a superoscillatory wave function would still produce a diffraction pattern from the open slit, but the pattern might exhibit unusual features compared to the ordinary case. Given the ability of superoscillatory functions to localize energy and oscillate at frequencies higher than their Fourier components would suggest, the diffraction pattern might include unexpectedly sharp features or localized intensity variations.
	The absence of a second slit eliminates the traditional form of interference between paths through different slits, but the superoscillatory nature of the wave function might still result in complex patterns due to the internal interference of its high-frequency components.

	In particular, in the limit of $n\rightarrow \infty$, for the case of both slit opened, the interference intensity on the screen is 
	\begin{equation}
		\label{Piccola1}
		I_{s}=\Big|\int_{\gamma_{A,B}} \mathcal{D}x[t]e^{i\mathcal{S}_{s}[x[t]]/\hbar}\Big|^{2},\,\,\,\, \mathcal{S}_{s}=\int \frac{dt}{a} \frac{a^2 p^2}{2m}=a \mathcal{S}_{o}\, , 
	\end{equation}
	where the integral is done on all over paths passing through the $A,B$ slits, $\mathcal{S}_{s,o}$ the superoscillation and ordinary action respectively. 
	This result suggests that in asymptotic regime of superoscillations, one obtain faster phase oscillations 
	from the $a$ factor rather than the ordinary case. 
	In particular, while the kinetic energy is amplified of a $a^{2}$ factor, the time is reduced as $a^{-1}$
	since, in non-relativistic regime $dt=dx/v$ with $v=dE/dp\rightarrow av$. 
	Such a result can be also re-interpreted as an effective reduced Planck constant $\hbar/a$. 
	On the other hand, at fix-n, the interference intensity can have a much more complicated profile 
	depending on the $n,a,k$ parameters with respect to characteristic interdistances. 
	
	In the framework of field theories, the implications of superoscillations could extend to how we understand the propagation and interactions of fields at quantum scales. Field theories generalize the concept of the quantum amplitude over space and time, incorporating the sum over histories approach to account for the vast array of possible field configurations. 
	In the next sections, we will move to a formulation of superoscillations in field theory, from canonical to path integral quantization.

	\section{Superoscillations in Quantum Field Theory}
	
	\subsection{Fields and Second Quantization}
	
	Let us now generalize the superoscillations to Spin-0 Relativistic Field Theories \footnote{In this section, we work in units $c=\hbar=1$. We work in the 4-dimensional Minkowski spacetime with the signature convention $(1,-1,-1,-1)$ on metric.}.
	As it is known, the scalar fields satisfy the 
	Klein-Gordon equation  
	\begin{equation}
		\label{eqq}
		(\Box+m^{2})\phi=0\, ,
	\end{equation}
	where $m$ is the mass of the spin 0 particle. 
	The two super-oscillating solutions at fixed four-momentum $p=(E,{\bf p})$ are 
	\begin{equation}
		\label{Tiem}
		\phi_{\pm}=b^{\pm}\sum_{k=0}^{n}c_{k}(n,a)e^{\pm i\left(1-\frac{2 k}{n}\right)(Et-{\bf p}\cdot {\bf x})}
	\end{equation}
	where $b^{\pm}\equiv b^{\pm}({\bf p})$.
	Such a solution is explicitly compatible with Lorentz invariance. 
	Indeed, in the limit of $n\rightarrow \infty$,  a-deformed relativistic plane waves are obtained 
	as 
	\begin{equation}
		\rightarrow b^{\pm}e^{\pm ia(Et-{\bf p}\cdot {\bf x})}\, . 
	\end{equation}

	A general solution, as a linear combination of the two above, and integrating on all possible four-momenta, reads as follows:
	\begin{equation}
		\label{sol}
		\phi=\int d^{4}p \mathcal{R}\sum_{k}\{c_{k}(n;a)(b^{-}({\bf p})e^{-i\left(1-\frac{2 k}{n}\right)px}+b^{+}({\bf p})e^{i\left(1-\frac{2 k}{n}\right)p\bf x})  \}\, .
	\end{equation}
	We demand
	that $(d^{4}p\mathcal{R})$ is Lorentz invariant, which means that the $\phi$ field 
	propagates in the light-cone and the energy is always positive.
	From the KG equation 
	\begin{equation}
		\label{DrivenD}
		(\Box+m^{2})\phi=0\rightarrow -\sum_{k=0}^{n}c_{k}(n;a)\left[\left(1-\frac{2 k}{n}\right)^2p^2-m^2\right]e^{i\left(1-\frac{2 k}{n}\right)px}=0\, ,
	\end{equation}
	we have the constrains 
	\begin{equation}
		\label{const}
		p^2\left(1-\frac{2 k}{n}\right)^2-m^2=0\rightarrow \delta\left(p^2\left(1-\frac{2 k}{n}\right)^2-m^2\right)\, .
	\end{equation}
	Thus, fixing k, we can impose the light cone constrains for each k-mode as
	\begin{equation}
		\label{kakaklaas}
		\theta(E)\delta\left(p^2\left(1-\frac{2 k}{n}\right)^2-m^2\right) \rightarrow \frac{1}{2(1-2k/n)^2 E}\delta(E-E_{+,k,n})\, , 
	\end{equation}
	where $E_{+,k,n}=\sqrt{|{\bf p}|^2+m^2/(1-2k/n)^2}$ is the positive zero of Eq.\ref{const}.

	Therefore, the classical solution can be written as 
	\begin{equation}
		\label{clll}
		\phi=\sum_{k=0}^{n}c_k\int \frac{d^{3}p}{(2\pi)^3\, 2\left(1-\frac{2 k}{n}\right)^2 E_{k}}\left(b^{-}({\bf p})e^{-i\left(1-\frac{2 k}{n}\right)px}+b^{+}({\bf p})e^{i\left(1-\frac{2 k}{n}\right)px}\right)\, . 
	\end{equation}
	
	Let us note that the mode $k=n/2$ corresponds 
	to an effective mass that is divergent 
	with the plane wave $e^{i(1-2k/n)px}=1$. 
	Such a k-mode is non-dynamical 
	and we will exclude it from the quantization. 
	
	The conjugate momentum of the scalar field is 
	\begin{equation}
		\label{clll}
		\pi=\sum_{k=0}^{n}c_k\int \frac{d^{3}\bf{p}}{(2\pi)^3}\frac{ i}{ 2\left(1-\frac{2 k}{n}\right)}\left(-b^{-}({\bf p})e^{-i\left(1-\frac{2 k}{n}\right)px}+b^{+}({\bf p})e^{i\left(1-\frac{2 k}{n}\right)px}\right)\, . 
	\end{equation}

	We can promote the classical field in Eq.\ref{clll} to a quantum field operator $\phi\rightarrow \hat{\phi}$ imposing the first quantization condition 
	\begin{equation}
		\label{CQ}
		[\hat{\phi}({\bf x}),\hat{\pi}({\bf y})]=i \delta^{3}({\bf x}-{\bf y})\, .
	\end{equation}
	where $\hat{\pi}$ is the conjugate momentum field operator. 
	The $b^{\pm}$ are promoted to k-dependent creation/annihilation operators 
	compatible with Eq.\ref{CQ}.
	The corresponding quantum fields are 
	\begin{equation}
		\label{QF1}
		\hat{\phi}=\sum_k\int \frac{d^{3}\bf p}{(2\pi)^3}\frac{1}{ \sqrt{2(n+1)E_k}}\left(\hat{b}({\bf p},k,n,a)e^{-i(1-\frac{2k}{n}) px}+\hat{b}^{\dagger}({\bf p},k,n,a)e^{i(1-\frac{2k}{n}) px}\right)\, ,
	\end{equation}
	\begin{equation}
		\label{QF2}
		\hat{\pi}=-i\sum_k \int \frac{d^{3}\bf p}{(2\pi)^3}\frac{(1-\frac{2k}{n})\sqrt{E_k}}{\sqrt{2 (n+1)}}\left(\hat{b}({\bf p},k,n,a)e^{-i(1-\frac{2k}{n}) px}-\hat{b}^{\dagger}({\bf p},k,n,a)e^{i(1-\frac{2k}{n}) px}\right)\, ,
	\end{equation}
	with 
	\begin{equation}
		\label{caop}
		[\hat{b}({\bf p},k,n,a),\hat{b}^{\dagger}({\bf p'},k',n,a)]=(2\pi)^3\delta({\bf p}-{\bf p'})\delta_{k,k'}\, ,
	\end{equation}
	where 
	\begin{equation}
		\label{defoe}
		\hat{b}({\bf p},k,n,a)=\frac{c_{k}\sqrt{n+1}}{(1-2k/n)^2 \sqrt{2 E_k}}\hat{b}^{-}({\bf p})\,,
	\end{equation}
	\begin{equation}
		\label{defoe2}
		\hat{b}^{\dagger}({\bf p},k,n,a)=\frac{c_{k}\sqrt{n+1}}{(1-2k/n)^2 \sqrt{2 E_k}} \hat{b}^{+}({\bf p})\,,
	\end{equation}
	for $k\neq n/2$.
	These equations imply the the quantum superoscillating fields 
	create and destroy simultaneously $n$ of the k-modes.

	In the limit of $n\rightarrow \infty$, the fields can be written as 
	\begin{equation}
		\label{PHIA}
		\phi=\int \frac{d^{3}p}{(2\pi)^3 2a^{2}E_{a}}\left(b^{-}({\bf p})e^{-iapx}+b^{+}({\bf p})e^{iapx}\right)\, ,
	\end{equation}
	where 
	\begin{equation}
		\label{tilde}
		E_{a}=\sqrt{{\bf p} ^2+\frac{m^2}{a^2}} .
	\end{equation}
	Let us quantize Eq.\ref{PHIA} as  
	
	\begin{equation}
		\label{PHIA2}
		\hat{\phi}=\int \frac{d^{3} \bf p}{(2\pi)^3}\frac{1}{\sqrt{ 2E_{a}}}[\hat{b}({\bf p},a)e^{-iapx}+\hat{b}^{\dagger}({\bf p},a)e^{iapx}]\, ,
	\end{equation}
	where
	\begin{equation}
		\label{lllalalaa}
		[\hat{b}({\bf p},a),\hat{b}^{\dagger}({\bf p'},a)]=(2\pi)^{3}\delta({\bf p}-{\bf p'})\, ,
	\end{equation}
	where 
	\begin{equation}
		\label{kiopn}
		\hat{b}({\bf p},a)=\frac{1}{a^2 \sqrt{2E_{a}}}\hat{b}^{-}({\bf p})\, , 
	\end{equation}
	\begin{equation}
		\label{kiopn}
		\hat{b}^{\dagger}({\bf p},a)=\frac{1}{a^2 \sqrt{2E_{a}}}\hat{b}^{+}({\bf p})\, .
	\end{equation}
	This can be interpreted as a second quantization procedure around the asymptotic limit.
	It is worth to note that, 
	in the limit of $n\rightarrow \infty$, 
	the creation/annihilation operators are not k-mode dependent anymore. 
	
	The Hamiltonian for the superoscillating scalar field in the large $n$ limit can be obtained as
	\begin{equation}
		\begin{aligned}		
			\hat{H}	& =\int d^3 x\left(\frac{1}{2} \hat{\pi}^2(t, \boldsymbol{x})+\frac{1}{2} \nabla^2 \hat{\phi}(t, \boldsymbol{x})+\frac{1}{2} m^2 \hat{\phi}^2\right) \\
			& 	=\left.\int \frac{d^3 \boldsymbol{p}}{(2 \pi)^3} a E\left(\hat{b}^{\dagger}(\boldsymbol{p}, a) \hat{b}(\mathbf{p}, a)+\frac{1}{2}(2 \pi)^3 \delta(\mathbf{p}-\mathbf{p})\right)\right|_{E=E_{a}}
		\end{aligned}
	\end{equation}
	The $\delta (0) $ term is infinite that present the vacuum energy. As for the ordinary harmonic oscillator, in larger n-limit, the energy levels of the superoscillations are  $~(N+\frac{1}{2})aE,N=0,1,2,...$.

	\subsection{Extended Fock Space}
	The quantum superoscillating field operator can act only on a  bandwidth limited Fock space. 
	In non-relativistic case one can consider a wave function $\phi({\bf x},t)$ in a bandwidth large as $2\omega_{max}$
	with $\omega_{max}$ the maximal Fourier transform frequency. 
	In particular, we can consider wave functions which are squared integrable in $[-\omega_{max},\omega_{max}]$,
	i.e. a Hilbert space that is 
	$$H_{\omega_{max}}=L^{2}(-\omega_{max},\omega_{max})$$
	equipped with scalar product as
	$$\langle \varphi_{1}|\varphi_{2}\rangle=\int_{-\omega_{max}}^{\omega_{max}}d\omega \tilde{\varphi}_{1}^{*}(\omega)\tilde{\varphi}_{2}(\omega)\, .$$
	
	In case of quantum field theories, 
	we generalise these definitions to a bandwidth limited Fock space as
	\begin{equation}
		\label{FockSpa}
		F\equiv Exp_{\alpha}(H_{\omega_{max}}) = \oplus_{m=0}^{\infty}(H^{ m}_{\omega_{max}})_{\alpha},\,\,\, \alpha=asym, sym
	\end{equation}
	where $F$ is the Fock space, $H$ the Hilbert space, $\oplus$ is the direct orthogonal sum of the Hilbert spaces
	$(H^{\otimes 0})_{\alpha}={\bf C}^{1}$,  $(H^{\otimes 1})_{\alpha}=H$, 
	$(H^{\otimes m})_{\alpha}$ $m>1$ are the symmetrized $\alpha=sym$ and antisymmetrized $\alpha=asym$
	tensor power of $H$.
	In other words, we construct the Fock spaces in the same way of ordinary quantum field theories
	\cite{Fock1,Fock2} but starting from bandwidth limited Hilbert spaces.
	
	Eq.\ref{FockSpa} is interpreted as the direct sum of m-Hilbert space, 
	where $m=0$ corresponds to the vacuum state $|0\rangle$
	which is one-dimensional space spanned by the vacuum vector, 
	$m=1$ to single-particle space $|1\rangle$;
	for $m$-Particle space the corresponding $|m\rangle$; in the case of bosons, these spaces are symmetric $\alpha=sym$ for particle exchange, 
	while for fermions, they are antisymmetric $\alpha=asym$ in 
	accordance with the Pauli exclusion principle.

	In particular, we are interested to Fock spaces that are constructed 
	on the basis of $L_{2}$ band-limited Hilbert spaces.
	In this case the Fock space is formed 
	by infinite sequences of the form 
	\begin{equation}
		\label{SequenceF}
		F=\{f_{0},f_{1},...,f_{m},...\}
	\end{equation}
	where 
	\begin{equation}
		\label{fzfufd}
		f_{0}\in {\bf C}^{1},\,\,\, f_{1}\in L_{2}(-\omega_{max},\omega_{max})\, , 
	\end{equation}
	\begin{equation}
		\label{fnnn}
		f_{m}\in L_{2}^{\alpha}(R^{m}_{\nu})_{[-\omega_{max},\omega_{max}]}\,,\,\,\, m=2,3,...,\,\,\, \nu=1,2,...
	\end{equation}
	and the product defined
	as 
	\begin{equation}
		\label{proddF}
		(F,G)=f_{0}\bar{g}_{0}+\sum_{m=1}^{\infty}(f_{m},g_{m})_{L_{2}^{\alpha}(R^{m},d_{\nu})_{[-\omega_{max},\omega_{max}}]}\, . 
	\end{equation}
	The notation $L_{2}^{\alpha}(\mathbb{R}^{m}, d_{\nu})_{[-\omega_{max},\omega_{max}]}$ in the context of Fock spaces refers to a space of square-integrable functions over $\mathbb{R}^{m}$ restricted to the bandwidth limit $[-\omega_{max},\omega_{max}]$, with a focus on either symmetrization or antisymmetrization, depending on the type of particles being described. Here, \(\alpha\) indicates whether the space is symmetrized for bosons or antisymmetrized for fermions, \(m\) represents the dimensionality of the space, and \(\nu\) signifies different measures or weightings applied within the space.
	The dimension \(m\) indicates the number of particles or spatial dimensions, and the measure \(d_{\nu}\) can represent physical properties such as mass distribution.

	The asymmetrized and symmetrized states 
	are the tensor products of the $f_{i}$
	as 
	\begin{equation}
		\label{tensorpro}
		(f_{1}\otimes ... \otimes f_{m})_{\alpha}\in F^{\alpha}(H),\,\,\, \alpha=s,a
	\end{equation}
	while the annihilation operators act as 
	$$b_{\alpha}(f)(f_{1}\otimes ...\otimes f_{m})_{\alpha}=\sum_{j=1}^{m}(-1)^{g_{\alpha}(j)}(f_{j},f)$$
	\begin{equation}
		\label{creaaan}
		\times (f_{1}\otimes ... \otimes f_{j-1}\otimes f_{j+1}\otimes ... \otimes f_{m})_{\alpha}\, , 
	\end{equation}
	with 
	\begin{equation}
		\label{AAa}
		b_{\alpha}(f)\Omega=0,\,\,\, \Omega=(1,0,...,0)\,, 
	\end{equation}
	where $\Omega$ represent the vacuum state. 
	The creation operator works as 
	\begin{equation}
		\label{actt}
		b_{\alpha}^{\dagger}(f)(f_{1}\otimes ... \otimes f_{m})=f\otimes (f_{1}\otimes ... \otimes f_{m})
	\end{equation}
	and 
	\begin{equation}
		\label{adag}
		b_{\alpha}^{\dagger}(f)\Omega=f\, . 
	\end{equation}

	The creation/annihilation operators not only can be of ordinary kind
	but they can also be related to superoscillations. 
	
	In particular a superoscillation is created with 
	the simultaneous application of $k=0,...,n$, $\hat{b}^{\dagger}(f_{k,n,a})$
	on the Fock space
	\begin{equation}
		\label{simbb}
		\prod_{k=0}^{n}\hat{b}^{\dagger}(f_{k,n,a})(f_{1}\otimes ... \otimes f_{m})=f(n,a)\otimes (f_{1}\otimes ... \otimes f_{m})\, ,
	\end{equation}
	where $f(n,a)$ is the superoscillating state. 
	
	On the other hand, we wish to include interactions of superoscillating fields
	also with ordinary fields which are not living in the band width limited Fock space. 
	In other words, we want to consider interaction terms 
	that involve not only fields acting on the bandwidth limited space
	but also standard ones considered in no-limited spaces. 
	This is in principle possible as sum of the two different Fock spaces:
	\begin{equation}
		\label{FSS}
		F_{U}\oplus F_{L}=Exp_{\alpha}(H_{U})\oplus Exp_{\alpha}(H_{L})
	\end{equation}
	where $U,L$ denote unlimited and limited bandwidth respectively. In next section, such a notation will be also useful for introducing the path integral formulation for superoscillating fields interacting with ordinary ones.

	\subsection{Path integral}

	Alternatively to the canonical second quantization studied above, 
	one can also consider other ways such as the path integral. 
	In this section, we will discuss the spin 0 field theory. 
	The path integral for scalar fields reads as 
	\begin{equation}
		\label{ZZ}
		Z[J_{U},J_{L}]=\int \mathcal{D}\phi_{U} \mathcal{D}\phi_{L}e^{i \mathcal{S}[\phi_{U},\phi_{L}]+J_{U,L}\phi_{U,L}}\, ,
\end{equation}
where $\mathcal{S}$ is the action which is a functional of $\phi_{U}$ and $\phi_{L}$.
In the path integral defined in Eq.\ref{ZZ}, we integrate on both fields acting on unlimited (U) and limited (L) bandwidth Fock spaces, defined in the previous section.
The $\phi_{L}$ includes superoscillating quantum fields. 

For example, it is possible to introduce a lagrangian density as 
\begin{equation}
	\label{LLL}
	\mathcal{L}[\phi_{U},\phi_{L}]=\frac{1}{2}m^{2}\phi_{U}^{2}+\frac{1}{2}m^{2}\phi_{L}^{2}+\frac{1}{4!}\lambda(\phi_{U}+\phi_{L})^{4}\
\end{equation}
mixing operators acting on limited and unlimited Fock spaces of states. 

Let us perform a double expansion with respect to sources of $U,L$ fields, $J_{U},J_{L}$ respectively, 
as follows: 
\begin{equation}
	\label{ZZJJ}
	Z[J_{U},J_{L}]=Z_0\sum_{n=0}^{\infty}\frac{i^{n}}{n!}\int dx_{1}...dx_{n}J_{U,L}(x_{1})...J_{U,L}(x_{n})G^{(n)}(x_{1},...,x_{n})\, 
\end{equation}
with $Z_0\equiv Z(0,0)$ and 
\begin{equation}
	\label{Gnnn}
	G^{n}(x_{1},...,x_{n})=\frac{1}{Z(0)}\int \mathcal{D}\phi_{U,L} e^{i\mathcal{S}}\phi_{U,L}(x_{1})...\phi_{U,L}(x_{n})\, .
\end{equation}

The 2-point Green's function corresponds to 
\begin{equation}
	\label{G2}
	G(x_{1},x_{2})=\frac{1}{Z(0)}\int \mathcal{D}\phi_{U,L} e^{i\mathcal{S}}\phi_{U,L}(x_{1})\phi_{U,L}(x_{2})\, . 
\end{equation}
In the limit of $\lambda\rightarrow 0$ or more in general in case of 
all couplings to zero except the mass,
Eq.\ref{G2} corresponds to the propagator. 
This reconfirms that the mathematical structure of the propagator in case of superoscillating fields
is the same of ordinary fields.

The 4-point Green's function reads as 
\begin{equation}
	\label{GGG}
	G(x_{1},x_{2},x_{3},x_{4})=\frac{1}{Z(0)}\int \mathcal{D}\phi_{U,L} e^{i\mathcal{S}}\phi_{U,L}(x_{1})\phi_{U,L}(x_{2})\phi_{U,L}(x_{3})\phi_{U,L}(x_{4})\, , 
\end{equation}

It is worth to remind that in general superoscillations can have different function profiles 
than Eq.\ref{super} (see Ref.\cite{Review} for an overview on these aspects). 
Thus, one may wonder how many superoscillating functions should be included in the path integral.
Concerning this issue, the quantization procedure is {\it system dependent}: 
a certain physical system allows for a specific superoscillation, rather than others,
included in the path integral. 
We can dub it {\it Principle of System Cluster Inclusion}.

From the path integral, we can directly formulate a definition of {\bf S}-matrix with an explicit separation of 
unlimited from limited bandwidth fields.
Indeed, the field in the path integral can be rewritten as 
the addition of interacting fields $\phi_{U,L}$ and a free asymptotic field 
$\phi_{U,L\, \infty}$ as 
\begin{equation}
	\label{asym}
	\phi_{U} \rightarrow \phi_{U}+\phi_{U\,\infty},\,\,\,\phi_{L} \rightarrow \phi_{L}+\phi_{L\,\infty}\, ,
\end{equation}
where 
\begin{equation}
	\label{satisfyy}
	(\Box +m^{2})\phi_{U,L\,\, \infty}=0\, . 
\end{equation}
The S-matrix can be expressed as 
$${\bf S}_{0}[J_{U},J_{L}]=\int \mathcal{D}\phi_{U}\mathcal{D}\phi_{L}{\rm exp}\big\{-i\int \frac{1}{2}\phi_{U,L}[\Box+m^{2}]\phi_{U,L}$$
\begin{equation}
	\label{expressS}
	-J_{U,L}\phi_{U,L}-J_{U,L}\phi_{U,L\, \infty}\}=e^{i \phi_{U\,\infty} J_{U}}e^{i \phi_{L\, \infty}J_{L}}W_{0}[J_{U},J_{L}]\, .
\end{equation}
Eq.\ref{expressS} holds also including interactions. 
In case of interactions, ${\bf S}[J_{U},S_{L}]$ can be expanded in power series as 
\begin{equation}
	\label{expandedSSS}
	{\bf S}[J_{U},J_{L}]=\sum_{n=0}^{\infty}\sum_{l=0}^{\infty}\frac{1}{n!l!}[i\phi_{U,\infty}J_{U}]^{n}[i\phi_{L,\infty}J_{L}]^{l}W[J_{U},J_{L}]\, .
\end{equation}

In the case L-fields do not have any interaction portals with ordinary fields in the unlimited Fock space,
Eq.\ref{expandedSSS} can be completely factorized as ${\bf S}={\bf S}_{U}{\bf S}_{L}$.

The path integral approach to correlators can be compared to ones from 
the time evolution operator. 
As in ordinary fields, in superoscillations the unitary time evolution operator satisfies 
\begin{equation}
	\label{UU}
	i\frac{\partial}{\partial t}\mathcal{U}(t,t_{0})=H_{I}(t)\mathcal{U}(t,t_{0})\, , 
\end{equation}
where $H_{I}(t)$ is the interaction Hamiltonian defined for free asymptotic fields. 
Then the formal solution to this equation is 
$$\mathcal{U}(t,-\infty)=T\, {\rm exp}\big(-i \int_{-\infty}^{t}dt_{1}H_{I}(t_{1})\big)$$
\begin{equation}
	\label{foo}
	=T\, {\rm exp}\big(-i \int_{-\infty}^{t}dt_{1}\int d^{3}{\bf x}_{1}\mathcal{H}_{I}({\bf x}_{1},t_{1})\big)
\end{equation}
where $T$ is the time order operator. 
The Green's function corresponds to 
$$G(x_{1},...,x_{n})=\sum_{m=0}^{\infty}\frac{(-1)^{m}}{m!}\int_{-\infty}^{\infty}d^{4}y_{1}...d^{4}y_{m}$$
\begin{equation}
	\label{Gree}
	\times \langle 0|T\Big[ \phi_{U,L}(x_{1})...\phi_{U,L}(x_{n})\mathcal{H}_{I}(y_{1})...\mathcal{H}_{I}(y_{m})\Big]|0\rangle_{{\bf c}}
\end{equation}
with ${\bf c}$ referring to connected diagrams only. 
Following this approach it is straightforward to prove that the Wick theorem is also valid for superoscillating fields. 
For example, in case of $\lambda \phi^4$ theory,
the four-point function can be expanded to the first order as 
$$G(x_{1},...,x_{4})=-\frac{i\lambda}{4!}\int d^{4}y \langle 0|T\Big[\phi_{U,L}(x_{1})...\phi_{U,L}(x_{4}) \Big]|0\rangle+...$$
\begin{equation}
	\label{G}
	=(-i\lambda) \int d^{4}y \prod_{i=1}^{4}[i\Delta_{F}(x_{i}-y)]+...
\end{equation}
where we used the Wick theorem and the fact the propagator for ordinary and superoscillating fields is the same
in mathematical definition but with different inputs as remarked above. 

It is worth to comment on Feynman's rules in case of superoscillations for $n>>1$.
As a simple case, we consider the scalar field theory $\lambda \phi^4$. 
Feynman's rules read as follows:
(i) draw all possible connected diagrams which have different topologies
with $n$ external lines, including loops;
(ii) associate a propagator $i\Delta_{F}(p)=\frac{i}{a^2 p^{2}-m^{2}+i\epsilon}$ for each internal line; 
(iii) associate $-i\lambda$ for each interaction vertex; 
(iV) associate an integration factor for each internal momentum of loops 
as $\int \frac{d^{4}p}{(2\pi)^{4}}$; (V) each graph has to be divided by an overall symmetry  factor
which corresponds to the number of permutation of vertices and internal lines at fixed external legs; 
(VI) momentum is conserved at each interaction vertex.
Indeed, Feynman's rules are basically unchanged with respect to the ordinary case.
Nevertheless what is changing is the kinematic in-put in the scattering amplitude. 
We will comment on implications of it in the next section concerning applications in scatterings.

\vspace{0.2cm}
\subsection{Fermion fields}

As it is well known spin 1/2 fields are fermions satisfying the 
Dirac equation 
\begin{equation}
	\label{DiracF}
	(i\gamma^{\mu}\partial_{\mu}-m)\psi=0\, , 
\end{equation}
with the $\gamma$-matrices satisfying the Clifford algebra
$\{\gamma^{\mu},\gamma^{\nu}\}=2\eta^{\mu\nu}$ with $\{...\}$ the anti-commutator. 
Here we assume the absence of any sources. 
A specular equation can be written for the $\bar{\psi}=\psi^{\dagger}\gamma^{0}$ field 
as 
\begin{equation}
	\label{DiracFRL}
	\bar{\psi}(i\gamma^{\mu}\partial_{\mu}+m)=0\, . 
\end{equation}
where here $\partial_{\mu}$ is thought as a differential operator acting on the left side of the expression. 

As it is known, both equations can be derived from the lagrangian density 
\begin{equation}
	\label{LDir}
	\mathcal{L}(\psi,\bar{\psi})=\bar{\psi}(i\gamma^{\mu}\partial_{\mu}-m) \psi\, . 
\end{equation}
Variations of this lagrangian with respect to $\psi,\bar{\psi}$ will generate Eq.\ref{DiracF} and Eq.\ref{DiracFRL}. 

The energy-momentum reads as 
\begin{equation}
	\left(1-\frac{2 k}{n}\right)^2 p^2-m^2=0\, .
\end{equation}
Thus the covariant momentum integral is the same as scalar case, 
\begin{equation}
	\int d^4 p \mathcal{R}=\int d^3 \boldsymbol{p} \frac{1}{2\left(1-\frac{2 k}{n}\right)^2 E(k, \boldsymbol{p})}\, ,
\end{equation}
where $E(k, \boldsymbol{p})=\sqrt{\frac{m^2}{\left(1-\frac{2 k}{n}\right)^2}+\boldsymbol{p}^2}$.

From the two Equations of Fields, we find superoscillating solutions as 
\begin{equation}
	\begin{aligned}
		\psi(t, \boldsymbol{x})  =&\sum_{s=1,2} \sum_{k=0}^n c_k(n, a) \int \frac{d^3 \boldsymbol{p}}{(2 \pi)^3} \frac{1}{2\left(1-\frac{2 k}{n}\right)^2 E(k, \boldsymbol{p})} \\
		&	\times \left(b^s(\boldsymbol{p}) u^s(p) e^{-i\left(1-\frac{2 k}{n}\right)px}+d^{s *}(\boldsymbol{p}) v^s(p) e^{i\left(1-\frac{2 k}{n}\right)px}\right) \\
		\bar{\psi}(t, \boldsymbol{x}) =&\sum_{s=1,2} \sum_{k=0}^n c_k(n, a) \int \frac{d^3 \boldsymbol{p}}{(2 \pi)^3} \frac{1}{2\left(1-\frac{2 k}{n}\right)^2 E(k, \boldsymbol{p})}\\	
		& \times \left(d^s(\boldsymbol{p}) \bar{v}^s(p) e^{-i\left(1-\frac{2 k}{n}\right)px}+b^{s *}(\boldsymbol{p}) \bar{u}^s(p) e^{i\left(1-\frac{2 k}{n}\right)px}\right)
	\end{aligned}
	\label{hmmN}
\end{equation}

where $u,v$ are the spinors as in standard fermion field theory.

We promote the classical field to quantum  operator and normalize it as follows
\begin{equation}
	\begin{aligned}
		\hat{\psi}(t, \boldsymbol{x})=&\sum_{s=1,2} \sum_{k=0}^n \int \frac{d^3 \boldsymbol{p}}{(2 \pi)^3} \frac{\sqrt{\left(1-\frac{2 k}{n}\right)}}{\sqrt{2 (n+1) E(k, \boldsymbol{p})}}\\
		& \times \left(\hat{b}^s(\boldsymbol{p}, k,n,a) u^s(p) e^{-i\left(1-\frac{2 k}{n}\right)px}+\hat{d}^{s \dagger}(\boldsymbol{p}, k,n,a) v^s(p) e^{i\left(1-\frac{2 k}{n}\right)px}\right)\, , \\
		\hat{\bar{\psi}}(t, \boldsymbol{x})=&\sum_{s=1,2} \sum_{k=0}^n \int \frac{d^3 \boldsymbol{p}}{(2 \pi)^3} \frac{\sqrt{\left(1-\frac{2 k}{n}\right)}}{\sqrt{2 (n+1) E(k, \boldsymbol{p})}}\\
		& \times \left(\hat{d}^s(\boldsymbol{p}, k,n,a) \bar{v}^s(p) e^{-i\left(1-\frac{2 k}{n}\right)px}+\hat{b}^{s \dagger}(\boldsymbol{p}, k,n,a) \bar{u}^s(p) e^{i\left(1-\frac{2 k}{n}\right)px}\right)\, ,
	\end{aligned}
\end{equation}
with 
\begin{equation}
	\begin{aligned}
		\hat{b}^s(\boldsymbol{p}, k, n, a) & =\sqrt{n+1} \frac{c_k(n, a)}{\left(1-\frac{2 k}{n}\right)^{5 / 2} \sqrt{2 E_k}} \hat{b}^s(\boldsymbol{p})\, , \\
		\hat{b}^{s \dagger}(\boldsymbol{p}, k, n, a) & =\sqrt{n+1} \frac{c_k(n, a)}{\left(1-\frac{2 k}{n}\right)^{5 / 2} \sqrt{2 E_k}} \hat{b}^{s \dagger}(\boldsymbol{p})\, , \\
		\hat{d}^s(\boldsymbol{p}, k, n, a) & =\sqrt{n+1} \frac{c_k(n, a)}{\left(1-\frac{2 k}{n}\right)^{5 / 2} \sqrt{2 E_k}} \hat{d}^s(\boldsymbol{p})\, , \\
		\hat{d}^{s \dagger}(\boldsymbol{p}, k, n, a) & =\sqrt{n+1} \frac{c_k(n, a)}{\left(1-\frac{2 k}{n}\right)^{5 / 2} \sqrt{2 E_k}} \hat{d}^{s \dagger}(\boldsymbol{p})\, .
	\end{aligned}
\end{equation}

The conjugate momentum is related to $\hat{\psi}^{\dagger}$ 
while the quantization condition is written with the anti-commutator rather than the commutator: 
\begin{equation}
	\label{QCC}
	\hat{\pi}=\frac{\delta \mathcal{L}}{\delta \dot{\hat{\psi}}}=i \hat{\psi}^{\dagger}=i \hat{\bar{\psi}} \gamma^0  ,\,\,\,\,\{\hat{\psi}_{i}({\bf x},t),\hat{\psi}^{\dagger}_{j}({\bf y},t)\}=\delta_{ij}\delta^{(3)}({\bf x}-{\bf y})\, . 
\end{equation}

Substituting Eq.\ref{hmmN} and its Hermitian conjugate inside the quantization condition, 
we find for the the two set of creation/annihilation operators that 
\begin{equation}
	\label{bbdag}
	\{\hat{b}_{r}({\bf p},k,n,a),\hat{b}_{s}^{\dagger}({\bf p}',k',n,a)\}=(2 \pi )^3\delta_{rs}\delta_{kk'}\delta^{(3)}({\bf p}-{\bf p'})\, , 
\end{equation}
\begin{equation}
	\label{bbdag2}
	\{\hat{d}_{r}({\bf p},k,n,a),\hat{d}_{s}^{\dagger}({\bf p}',k',n,a)\}=(2 \pi)^3\delta_{rs}\delta_{kk'}\delta^{(3)}({\bf p}-{\bf p'})\, . 
\end{equation}

Concerning the physical interpretation of $\hat{\psi}$, 
$$\sum_{k=0}^{n} \hat{b}({\bf p},k,n,a)u(p)e^{-i(1-2k/n)px}$$
corresponds to the annihilation of $k=0,...,n$
modes for the $E>0$ fermion; 
$$\sum_{k=0}^{n}\hat{d}^{\dagger}({\bf p},k,n,a)v(p)e^{i(1-2k/n)px}$$
to creation of of $k=0,...,n$ modes for $E>0$ anti-fermions.

In the large $n$ limit, the  superoscillating Dirac field is 
\begin{equation}
	\begin{aligned}
		& \psi(t, \boldsymbol{x})=\sum_{s=1,2} \int \frac{d^3 \boldsymbol{p}}{(2 \pi)^3} \frac{1}{2 a^2 E_{a}( \boldsymbol{p})}\left(b^s(\boldsymbol{p}) u^s(p) e^{-i apx}+d^{s *}(\boldsymbol{p}) v^s(p) e^{i apx}\right)\, , \\
		& \bar{\psi}(t, \boldsymbol{x})=\sum_{s=1,2} \int \frac{d^3 \boldsymbol{p}}{(2 \pi)^3} \frac{1}{2 a^2 E_{a}( \boldsymbol{p})}\left(d^s(\boldsymbol{p}) \bar{v}^s(p) e^{-i apx}+b^{s *}(\boldsymbol{p}) \bar{u}^s(p) e^{i apx}\right)\, .
	\end{aligned}
\end{equation}
Accordingly, the operator field is given by 
\begin{equation}
	\begin{aligned}
		&\hat{\psi}(t, \boldsymbol{x})=\sum_{s=1,2} \int \frac{d^3 \boldsymbol{p}}{(2 \pi)^3} \frac{\sqrt{a}}{\sqrt{2 E_{a}( \boldsymbol{p})}}\left(\hat{b}^s(\boldsymbol{p}, a) u^s(p) e^{-i apx}+\hat{d}^{s \dagger}(\boldsymbol{p}, a) v^s(p) e^{i apx}\right)\, ,\\
		&\hat{\bar{\psi}}(t, \boldsymbol{x})=\sum_{s=1,2} \int \frac{d^3 \boldsymbol{p}}{(2 \pi)^3} \frac{\sqrt{a}}{\sqrt{2 E_{a}( \boldsymbol{p})}}\left(\hat{d}^s(\boldsymbol{p}, a) \bar{v}^s(p) e^{-i apx}+\hat{b}^{s \dagger}(\boldsymbol{p}, a) \bar{u}^s(p) e^{i apx}\right)\, ,
	\end{aligned}
\end{equation}
with the creation and annihilation operators defined as
\begin{equation}
	\begin{aligned}
		& \hat{b}^s(\boldsymbol{p}, a)=\frac{1}{a^{5 / 2} \sqrt{2 E_{a}}} \hat{b}^s(\boldsymbol{p}) \, , \\
		& \hat{b}^{s \dagger}(\boldsymbol{p}, a)=\frac{1}{a^{5 / 2} \sqrt{2 E_{a}}} \hat{b}^{s \dagger}(\boldsymbol{p}) \, , \\
		& \hat{d}^s(\boldsymbol{p}, a)=\frac{1}{a^{5 / 2} \sqrt{2 E_{a}}} \hat{d}^s(\boldsymbol{p}) \, , \\
		& \hat{d}^{s \dagger}(\boldsymbol{p}, a)=\frac{1}{a^{5 / 2} \sqrt{2 E_{a}}} \hat{d}^{s \dagger}(\boldsymbol{p}) \, .
	\end{aligned}
\end{equation}
The quantization conditions are $
\left\{\hat{b}^r(\boldsymbol{p},  a), \hat{b}^{s \dagger}\left(\boldsymbol{p}^{\prime},  a\right)\right\}=\left\{\hat{d}^r(\boldsymbol{p},  a), \hat{d}^{s \dagger}\left(\boldsymbol{p}^{\prime},  a\right)\right\}=(2 \pi)^3 \delta^{3}\left(\boldsymbol{p}-\boldsymbol{p}^{\prime}\right)  \delta^{r s}$ and $\left\{\hat{\psi}(t, \boldsymbol{x}), \hat{\psi}^{\dagger}(t, \boldsymbol{y})\right\}= \delta^{3}(\boldsymbol{x}-\boldsymbol{y})I_{4 \times 4}$.

It is worth to remark that, while both  "quantum" and "classical" bosonic superoscillating fields can be defined, 
superoscillating fermionic fields can only be "quantum" as a consequence of anti-commutativity.

\vspace{0.2cm}

\subsection{Spin 1 and QED}
Let us consider the case of Quantum Electrodynamics (QED), starting from the Lagrangian for Maxwell's theory without any sources 
as 
\begin{equation}
	\label{LLM}
	\mathcal{L}=-\frac{1}{4}F_{\mu\nu}F^{\mu\nu},\,\,\, F_{\mu\nu}=\partial_{\mu}A_{\nu}-\partial_{\nu}A_{\mu}
\end{equation}
and Euler-Lagrange equation $\partial_{\nu} F^{\mu\nu}=0$. 
As it is well known, Eq.\ref{LLM} is invariant under $U(1)$ gauge transformations $A_{\mu}(x)\rightarrow A_{\mu}(x)+\partial_{\mu}\lambda(x)$
corresponding to $F_{\mu\nu}\rightarrow F_{\mu\nu}$.

Let us fix our gauge as Coulomb-like $\nabla \cdot {\bf A}=0$. 
In this case the Equation of Motion for the vector field corresponds to 
\begin{equation}
	\label{AAeq}
	\Box  A^{i}(t, \boldsymbol{x})=\left(\partial_t^2-\partial_{\boldsymbol{x}}^2\right) A^i(t, \boldsymbol{x})=0\, 
\end{equation}
with $i=1,2,3$ spatial coordinate indices,
and such an equation has a superoscillating solution as follows:

\begin{equation}
	A^i(t, \boldsymbol{x})=\left.\sum_{k=0}^n c_k \int \frac{d^3 \boldsymbol{p}}{(2\pi)^3} \frac{1}{2 \left(1-\frac{2 k}{n}\right)^2 E} \sum_{r= \pm}\left\{b^r(\boldsymbol{p}) \epsilon_r^i(\boldsymbol{p}) e^{-i\left(1-\frac{2 k}{n}\right)px}+b^{r *}(\boldsymbol{p}) \epsilon_r^{* i}(\boldsymbol{p}) e^{i\left(1-\frac{2 k}{n}\right)p x}\right\}\right|_{E=|\boldsymbol{p}|}\, .
	\label{hmm}
\end{equation}
Here we used the covariant momentum integral for the massless spin-1 field by imposing the light-cone constrains $p^2=0$. To be consistent with the massive case, we include the k-factor as well, namely
\begin{equation}
	\int d^4 p \mathcal{R}=\int d^4 p \delta\left(\left(1-\frac{2 k}{n}\right)^2p^2\right)\theta (E) =\int \frac{d^3 \boldsymbol{p}}{(2 \pi)^3}  \frac{1}{2\left(1-\frac{2 k}{n}\right)^2E(\boldsymbol{p})}
\end{equation}
with $E(\boldsymbol{p})=|\boldsymbol{p}|$.

The polarization vectors satisfy the usual conditions 
${\bf \epsilon} \cdot {\bf p}=0$ and ${\bf \epsilon}_{r}\cdot {\bf \epsilon}_{s}=\delta_{rs}$. 

These solutions are compatible with the $U(1)$ Gauge invariance.

The conjugate momentum can be obtained by varying $\mathcal{L}=\frac{1}{2} \partial_t A_i \partial_t A_i-\frac{1}{2} \partial_j A_i \partial_j A_i+\cdots$ with respect to $\dot{A}_{i}$, namely
\begin{equation}
	\begin{aligned}
		&	\pi^i(t, \boldsymbol{x})=\partial_t A_i(t, \boldsymbol{x})=E^i\\
		&=i \sum_{k=0}^n c_k \int \frac{d^3 \boldsymbol{p}}{(2\pi)^3} \frac{1}{2 \left(1-\frac{2 k}{n}\right)} \sum_{r= \pm}\left\{-b^r(\boldsymbol{p}) \epsilon_r^i(\boldsymbol{p}) e^{-i\left(1-\frac{2 k}{n}\right)px}+b^{r *}(\boldsymbol{p}) \epsilon_r^{* i}(\boldsymbol{p}) e^{i\left(1-\frac{2 k}{n}\right)px}\right\} 
	\end{aligned}\,\,\, .
\end{equation}

We will now perform a second quantization of superoscillating fields 
promoting $A^i$ and $\pi^i$ to quantum operator fields:
\begin{equation}
	\begin{aligned}
		\hat{A}^i(t, \boldsymbol{x})  =&\sum_{k=0}^n \int \frac{d^3 \boldsymbol{p}}{(2 \pi)^3} \frac{1}{\sqrt{n+1} \sqrt{2 E}}\\
		& \times \sum_{r= \pm}\left\{\hat{b}^r(\mathbf{p}, k, n, a) \epsilon_r^i(\boldsymbol{p}) e^{-i\left(1-\frac{2 k}{n}\right)p x}+\hat{b}^{r \dagger}(\mathbf{p}, k, n, a) \epsilon_r^{* i}(\boldsymbol{p}) e^{i\left(1-\frac{2 k}{n}\right)p x}\right\} \, , \\
		\hat{\pi}^i(t, \boldsymbol{x}) =&i \sum_{k=0}^n \int \frac{d^3 \boldsymbol{p}}{(2 \pi)^3} \frac{\left(1-\frac{2 k}{n}\right)}{\sqrt{n+1} \sqrt{2}} \sqrt{E} \\
		&\times \sum_{r= \pm}\left\{-\hat{b}^r(\mathbf{p}, k, n, a) \epsilon_r^i(\boldsymbol{p}) e^{-i\left(1-\frac{2 k}{n}\right)px}+\hat{b}^{r \dagger}(\mathbf{p}, k, n, a) \epsilon_r^{* i}(\boldsymbol{p}) e^{i\left(1-\frac{2 k}{n}\right)px}\right\} \, ,
	\end{aligned}
\end{equation}
where 
\begin{equation}
	\begin{aligned}
		& \hat{b}^r(\mathbf{p}, k, n, a)=\sqrt{n+1} \frac{c_k(n, a)}{\left(1-\frac{2 k}{n}\right)^2\sqrt{2 E}} \hat{b}^r(\mathbf{p})\, , \\
		& \hat{b}^{r \dagger}(\mathbf{p}, k, n, a)=\sqrt{n+1} \frac{c_k(n, a)}{\left(1-\frac{2 k}{n}\right)^2\sqrt{2 E}} \hat{b}^{r \dagger}(\mathbf{p}) \,  .
	\end{aligned}
\end{equation}

We can consistently quantize the superoscillating fields 
imposing the canonical condition with constraints $\nabla \cdot{\hat {\bf A}}=\nabla \cdot \hat{{\bf E}}=0$: 
\begin{equation}
	\label{CQAE}
	[\hat{{\bf A}}_{i}({\bf x}),{\hat {\bf E}}_{j}({\bf y})]=i (\delta_{ij}-\partial_{i}\partial_{j}/\nabla^{2})\delta^{(3)}({\bf x}-{\bf y})
\end{equation}
compatible with 
\begin{equation}
	\label{COMP}
	[\hat{{\bf A}}_{i}({\bf x}),\hat{{\bf A}}_{j}({\bf y})]=[\hat{{\bf E}}_{i}({\bf x}),{\hat{\bf E}}_{j}({\bf y})]=0\, . 
\end{equation}
The creation/annihilation operator satisfies the 
standard algebra but with $(k,n,a)$-dependence: 
\begin{equation}
	\label{Standarddd}
	[\hat{b}^{r}({\bf p},k,n,a),\hat{b}^{s}({\bf q},k,n,a)]=0
\end{equation}
\begin{equation}
	\label{Standardddd1}
	[\hat{b}^{\dagger r}({\bf p},k,n,a),\hat{b}^{\dagger s}({\bf q},k,n,a)]=0\,
\end{equation}
\begin{equation}
	\label{Standard2}
	[\hat{b}^{r}({\bf p},k,n,a),\hat{b}^{\dagger s}({\bf q},k',n,a)]=(2\pi)^{3}\delta^{rs}\delta_{k,k'}\delta^{(3)}({\bf p}-{\bf q})\, . 
\end{equation}
These are obtained assuming the ordinary completeness condition 
$\sum_{r=1,2}\epsilon_{r}^{i}({\bf p})\epsilon_{r}^{i}({\bf p})=\delta^{ij}-p^{i}p^{j}/|{\bf p}|^{2}$.

In the large $n$ limit, the $\sum_{k=0}^n$ in superoscillation solution reduces to \begin{equation}
	A^i(t, \boldsymbol{x})=\left.\int \frac{d^3 \boldsymbol{p}}{(2 \pi)^3} \frac{1}{2 a^2 E} \sum_{r= \pm}\left\{b^r(\boldsymbol{p}) \epsilon_r^i(\boldsymbol{p}) e^{-i apx}+b^{r *}(\boldsymbol{p}) \epsilon_r^{* i}(\boldsymbol{p}) e^{i apx}\right\}\right|_{E=|\boldsymbol{p}|}\, .
\end{equation}
The second quantization is performed by promoting the classical field to the quantum operator field:
\begin{equation}
	\hat{A}^i(t, \boldsymbol{x})=\left.\int \frac{d^3 \boldsymbol{p}}{(2 \pi)^3} \frac{1}{\sqrt{2 E}} \sum_{r= \pm}\left\{\hat{b}^r(\boldsymbol{p},a) \epsilon_r^i(\boldsymbol{p}) e^{-i apx}+\hat{b}^{r \dagger}(\boldsymbol{p},a) \epsilon_r^{* i}(\boldsymbol{p}) e^{i apx}\right\}\right|_{E=|\boldsymbol{p}|}
\end{equation}
with the creation and annihilation operator given by  \begin{equation}
	\begin{aligned}
		& \hat{b}^r(\mathbf{p},a)=\frac{1}{a^2\sqrt{2 E}} \hat{b}^r(\mathbf{p}), \\
		& \hat{b}^{r \dagger}(\mathbf{p},a)=\frac{1}{ a^2 \sqrt{2 E}} \hat{b}^{r \dagger}(\mathbf{p}) .
	\end{aligned}
\end{equation}
The quantization conditions are $\left[\hat{b}^r(\mathbf{p},a), \hat{b}^{s \dagger}\left(\mathbf{p}^{\prime},a\right)\right]=(2 \pi)^3 \delta\left(\mathbf{p}-\mathbf{p}^{\prime}\right) \delta^{r s}$ and $\left[A^i(t, \boldsymbol{x}), \pi^j(t, \boldsymbol{y})\right]=i\left(\delta_{i j}-\frac{\partial_i \partial_j}{\partial^2}\right) \delta(\boldsymbol{x}-\boldsymbol{y})$. 

As an alternative but equivalent second quantization procedure see Appendix A.

\subsection{Spin 2 and Quantum Gravity}

Let us consider the linearized theory of gravity 
splitting the metric field tensor $g_{\mu\nu}$ into 
background $\bar{g}_{\mu\nu}$ and fluctuations $h_{\mu\nu}$:
\begin{equation}
	\label{ggbarh}
	g_{\mu\nu}(x)=\bar{g}_{\mu\nu}+h_{\mu\nu},\,\,\, g^{\mu\nu}=\bar{g}^{\mu\nu}-h^{\mu\nu}\, , 
\end{equation}
with $g^{\mu\nu}g_{\nu\lambda}=\delta^{\mu}_{\lambda}$. 
In flat space time $\bar{g}_{\mu\nu}$ reduces to 
$\eta_{\mu\nu}$ as understood. 

Considering the vacuum Einstein equations $R_{\mu\nu}=0$, 
perturbations around the background obey to the equation 
\begin{equation}
	\label{nablna}
	\bar{\nabla}^{\alpha}\bar{\nabla}_{\mu}h_{\alpha\nu}+\bar{\nabla}^{\alpha}\bar{\nabla}_{\nu}h_{\alpha\mu}-\bar{\nabla}_{\mu}\bar{\nabla}_{\nu}h_{\alpha}^{\alpha}-\bar{\nabla}^2 h_{\mu\nu}=0\, .
\end{equation}

Let us split the h-perturbation field isolating the transverse-traceless (TT) part 
as 
\begin{equation}
	\label{hmunuTT}
	h_{\mu\nu}=h_{\mu\nu}^{TT}+\bar{\nabla}_{(\mu}\zeta_{\nu)}+\frac{1}{D}\bar{g}_{\mu\nu}h\, , 
\end{equation}
where $D$ is the space-time dimension number and where the TT part satisfies the condition 
\begin{equation}
	\label{condition}
	\bar{\nabla}^{\mu}h_{\mu\nu}^{TT}=h_{\mu}^{\mu\, TT}=0\, . 
\end{equation}
The second and the third terms in Eq.\ref{hmunuTT}
are the gauge and trace parts respectively. {In addition, the time-component is chosen to vanish $h_{0\mu}^{TT}=0$.
	On flat space-time background, the TT-part satisfies the
	EoM as 
	\begin{equation}
		\label{nablaTTEoM}
		\bar{\nabla}^{2}h_{ij}^{TT}=0\, . 
	\end{equation}
	The superoscillation solution for GW in the transverse traceless gauge  is 
	\begin{equation}
		h_{i j}^{T T}(t, \boldsymbol{x})=\left.\sum_k c_k \int \frac{d^3 \boldsymbol{p}}{(2 \pi)^3} \frac{1}{2 \left(1-\frac{2 k}{n}\right)^2 E} \sum_{r= \pm}\left\{b^r(\boldsymbol{p}) \epsilon_{i j}^r(\boldsymbol{p}) e^{-i\left(1-\frac{2 k}{n}\right)px}+b^{r *}(\boldsymbol{p}) \epsilon_{i j}^{* r}(\boldsymbol{p}) e^{i\left(1-\frac{2 k}{n}\right)px}\right\}\right|_{E=|\boldsymbol{p}|}\, ,
	\end{equation}
	where $\epsilon_{i j}^r(\boldsymbol{p})$ is the polarization tensor
	and we fix $D=3+1$ number of dimensions. Only two polarized states ($r=\pm$) are physical due to the transverse gauge condition. Since $\mathcal{L}=\frac{1}{2} \partial^\mu h_{\rho \sigma} \partial_\mu h^{\rho \sigma}$ in the transverse traceless gauge, one can obtain the momentum conjugation $\pi_{i j}^{T T}(t, \boldsymbol{x})=\partial_t h_{i j}^{T T}(t, \boldsymbol{x})$, that is
	\begin{equation}
		\pi^{TT}_{i j}(t, \boldsymbol{x})=i \sum_k c_k \int d^3 \boldsymbol{p} \frac{1}{2 \left(1-\frac{2 k}{n}\right)} \sum_{r= \pm}\left\{-b^r(\boldsymbol{p}) \epsilon_{i j}^r(\boldsymbol{p}) e^{-i\left(1-\frac{2 k}{n}\right)px}+b^{r *}(\boldsymbol{p}) \epsilon_{i j}^{* r}(\boldsymbol{p}) e^{i\left(1-\frac{2 k}{n}\right)px}\right\}\, .
	\end{equation}

	We promote the coefficients $b^r(\boldsymbol{p})$ and its complex conjugation $b^{r *}(\boldsymbol{p})$ to creation/annihilation  operators and we normalise them as 
	\begin{equation}
		\begin{aligned}
			& \hat{b}^r(\mathbf{p}, k, n, a)=\sqrt{n+1} \frac{c_k(n, a)}{\left(1-\frac{2 k}{n}\right)^2\sqrt{2 E}} \hat{b}^r(\mathbf{p})\, , \\
			& \hat{b}^{r \dagger}(\mathbf{p}, k, n, a)=\sqrt{n+1} \frac{c_k(n, a)}{\left(1-\frac{2 k}{n}\right)^2\sqrt{2 E}} \hat{b}^{r \dagger}(\mathbf{p})\, ,
		\end{aligned}
	\end{equation}
	with 
	\begin{equation}
		\left[\hat{b}^r(\mathbf{p}, k, n, a), \hat{b}^{s \dagger}\left(\mathbf{p}^{\prime}, k^{\prime}, n, a\right)\right]=(2 \pi)^3 \delta\left(\mathbf{p}-\mathbf{p}^{\prime}\right) \delta_{k k^{\prime}} \delta^{r s}\, .
	\end{equation} 
	Accordingly, the classical GW fields and its momentum conjugation are 
	\begin{equation}
		\begin{aligned}
			\hat{h}_{i j}^{T T}(t, \boldsymbol{x}) & =\sum_k \int \frac{d^3 \boldsymbol{p}}{(2 \pi)^3} \frac{1}{ \sqrt{2(n+1) E}} \sum_{r= \pm}\left\{\hat{b}^r \epsilon_{i j}^r(\boldsymbol{p}) e^{-i\left(1-\frac{2 k}{n}\right)p x}+\hat{b}^{r \dagger} \epsilon_{i j}^{* r}(\boldsymbol{p}) e^{i\left(1-\frac{2 k}{n}\right)p x}\right\}, \\
			\hat{\pi}^{TT}_{i j}(t, \boldsymbol{x}) & =i \sum_k \int \frac{d^3 \boldsymbol{p}}{(2 \pi)^3} \frac{\left(1-\frac{2 k}{n}\right)\sqrt{E}}{ \sqrt{2(n+1)}}  \sum_{r= \pm}\left\{-\hat{b}^r \epsilon_{i j}^r(\boldsymbol{p}) e^{-i\left(1-\frac{2 k}{n}\right)px}+\hat{b}^{r \dagger} \epsilon_{i j}^{* r}(\boldsymbol{p}) e^{i\left(1-\frac{2 k}{n}\right)px}\right\}
		\end{aligned}
	\end{equation}
	which satisfy the quantization condition $\left[h_{\mu \nu}^{T T}(t, \boldsymbol{x}), \pi_{k l}(t, \boldsymbol{y})\right]=i\left(P_{i k} P_{j l}-\frac{1}{2} P_{i j} P_{k l}\right) \delta(\boldsymbol{x}-\boldsymbol{y})$, as expected. Here the projection operator is $P_{i j}=\delta_{i j}-\frac{p_i p_j}{\boldsymbol{p}^2}$ in momentum space and $P_{i j}=\delta_{i j}-\frac{\partial_i \partial_j}{\partial^2}$ in position space. To derive above formula, we have used the completeness condition on the polarization vector 
	$\sum_{r=\pm}\epsilon_{r}^{ij}({\bf p})\epsilon_{r}^{kl}({\bf p})=\left(P_{i k} P_{j l}-\frac{1}{2} P_{i j} P_{k l}\right) $. 
}

In the large $n$ limit, the superoscillation solution has a simpler form as follows
\begin{equation}
	h_{i j}^{T T}(t, \boldsymbol{x})=\left.  \int \frac{d^3 \boldsymbol{p}}{(2 \pi)^3} \frac{1}{2 a^2 E} \sum_{r= \pm}\left\{b^r(\boldsymbol{p}) \epsilon_{i j}^r(\boldsymbol{p}) e^{-iapx}+b^{r *}(\boldsymbol{p}) \epsilon_{i j}^{* r}(\boldsymbol{p}) e^{iapx}\right\}\right|_{E=|\boldsymbol{p}|}\, .
\end{equation}
The corresponding quantum field is 
\begin{equation}
	\hat{h}_{i j}^{T T}(t, \boldsymbol{x})=\left.  \int \frac{d^3 \boldsymbol{p}}{(2 \pi)^3} \frac{1}{\sqrt{2 E}} \sum_{r= \pm}\left\{\hat{b}^r(\boldsymbol{p},a) \epsilon_{i j}^r(\boldsymbol{p}) e^{-iapx}+\hat{b}^{r \dagger}(\boldsymbol{p},a) \epsilon_{i j}^{* r}(\boldsymbol{p}) e^{iapx}\right\}\right|_{E=|\boldsymbol{p}|}
\end{equation}
with  
\begin{equation}
	\begin{aligned}
		& \hat{b}^r(\mathbf{p},a)=\frac{1}{a^2\sqrt{2 E}} \hat{b}^r(\mathbf{p}) \, \\
		& \hat{b}^{r \dagger}(\mathbf{p},a)=\frac{1}{a^2\sqrt{2 E}} \hat{b}^{r \dagger}(\mathbf{p}),
	\end{aligned}
\end{equation}
with second quantization condition
\begin{equation}
	\left[\hat{b}^r(\mathbf{p},a), \hat{b}^{s \dagger}\left(\mathbf{p}^{\prime},a\right)\right]=(2 \pi)^3 \delta\left(\mathbf{p}-\mathbf{p}^{\prime}\right) \delta^{r s}\, .
\end{equation}
One can verify that the field and its momentum conjugation satisfy $\left[\hat{h}_{\mu \nu}^{T T}(t, \boldsymbol{x}), \hat{\pi}_{k l}^{T T}(t, \boldsymbol{y})\right]=i\left(P_{i k} P_{j l}-\frac{1}{2} P_{i j} P_{k l}\right) \delta(\boldsymbol{x}-\boldsymbol{y})$. 

\vspace{1cm}

Let us consider a superoscillating Gravitational Wave acting on two test particles. 

For simplicity, we suppose that the GW is a plane wave propagating along the 
z-direction as $h_{\mu}=\epsilon_{\mu}^{+}S(t-z)$ 
where $S(t-z)=\sum_{k=0}^{n}c_{k}(a;n)e^{i(1-2k/n)(Et-p_{z}z)}$
and $\epsilon^{+}=1$. 
The perturbed metric from GW is 
\begin{equation}
	\label{eeeds}
	ds^{2}=-dt^{2}+(1+S(t-z))dx^{2}+(1-S(t-z))dy^{2}+dz^2
\end{equation}
assuming $|S|<<1$. 
Let us assume that the two test-particles are disposed along the x-axis 
at interdistance of $D_{0}$ and initial four-velocity as $v_{A}^{\mu}=v_{B}^{\mu}=(1,0,0,0)$ (rest frame). 
Under this condition, the geodesic equation reduces to $dv^{\mu}/d\tau=0$.
In this situation the proper interdistance change 
as 
\begin{equation}
	\label{inter}
	\frac{D(t)-D_{0}}{D_{0}}\simeq \frac{1}{2}S(t)=\sum_{k=0}^{n}c_{k}(a;n)e^{i(1-2k/n)Et}\, . 
\end{equation}
In principle, such an effect can be tested in GW interferometers
such as aLIGO/VIRGO or space-based LISA.

\vspace{3cm}

\section{Applications}

\subsection{Scatterings}

As an implication of considering interaction terms among fields in limited and unlimited bandwidths,
scatterings among superoscillating and ordinary particles can be envisaged. 
The computations of scattering amplitudes can be performed using conventional 
Feynman's diagrams and rules with the  kinematic inputs of superoscillating particles. 
For example, we can consider an interaction vertex involving relativistic superoscillating and ordinary scalar fields 
as $\lambda_{1} \phi_{s}\phi_{o}^{3}$. Let us consider a physical set-up as in Fig.1: 
a superoscillating scalar field is emitted from a black box with a tiny hole compared to 
the box size similarly to how proposed in Refs.\cite{APR,APR2}. 
We consider a box with maximal Fourier frequency of $\omega_{max}$
while the $\phi_{s}$ is emitted, in asymptotic limit, with energy  $aE_{1}=a\hbar \omega_{max}$. 
The outgoing superoscillating particle scatters on an ordinary scalar field
producing two ordinary scalar fields $\phi_{o}(p_{3}),\phi_{o}(p_{4})$.
Here, we suppose that the ordinary and superoscillating particles have exactly the same charges and mass,
i.e. $\phi_{s}$ is considered as the superoscillating version of the field $\phi_{o}$.
Let us consider the scattering process in relativistic regime: the conservation of four momenta implies  $ap_1+p_2 =p_3+p_4$ rather than $p_1+p_2=p_3+p_4$,
where $p_{1}=(E_{1},{\bf p_{1}})$ is the four-momentum for the max Fourier energy related to the source Box. 

The fact that superoscillating fields scatter at higher energies than the max Fourier ones
permits to probe shorter distances as well as to test higher energy massive particles 
and interaction mediators than in ordinary cases. This argument can also be extended to Black Hole formation in high energy scatterings.
All in all, a Black Hole is formed 
when $E_{CM}\geq M_{Pl}$ and $b\leq R_S=2G_{N}E_{CM}^2$, where $E_{CM}$ is the Center of Mass energy, $M_{Pl}$ the Planck scale, $b$ the impact parameter, 
$R_{S}$ the BH radius, $G_{N}$ the Newton constant \cite{GS1,GS2}.
For a BH formation from scattering of two superoscillating particles, 
the kinematics will be revisited as 
$E_{CM}\geq M_{Pl}/a$ and $b \sqrt{-t} \sim 1/a$, where $t$ is the t-Mandelstam variable of the collision. This can also be  interpreted
as an effective rescaling of the Newton constant 
as $G_{N}\rightarrow a^{2}G_{N}$, which will also be relevant for following discussions in subsection III.C. 

Let us consider the case of resonances.
In resonant scatterings, 
the particle wave function typically corresponds to 
$\psi(t) \sim e^{-iMt-\Gamma t}$,
with a Fourier transformation 
$\phi(E)\sim \int dt' e^{iEt'}e^{-iMt'-\Gamma t'}\sim \frac{1}{(E-M)+i\Gamma}$,
with $M,\Gamma$ the mass and decay rate of the resonance particle. 
In case of a superoscillating wave function, 
the resonant wave function reads as 
\begin{equation}
	\label{con}
	\sum_{k}c_{k}e^{-i(1-2k/n)(M-i\Gamma)t}
	\rightarrow e^{-iaMt-a\Gamma t}\, . 
\end{equation}
Performing a Fourier transform on it, 
we obtain 
$\phi(E)\sim \frac{1}{(E-aM)+ia\Gamma}$.
This means that, in this case, the resonance peak is around $E\simeq aM$ rather than $M$. Moreover, the decay rate of superoscillating resonance will re-scale as the same $a$-factor as a a sharper peak. 
On the contrary, we can consider an ordinary resonance as obtained from the scattering of 
superoscillating fields. 
In this last case, the resonant wave function 
reads as 
\begin{equation}
	\label{READA}
	\psi(t)\sim e^{-iMt/a-\Gamma t/a}\rightarrow \phi(E)\sim \frac{1}{(aE-M)+i\Gamma}\, .
\end{equation}
The a-factor multiplying the energy $E$ is from the kinematics of external superoscillating particles participating in the scattering. 
This means that in principle, a scattering of superoscillating fields at a center of mass energy $E_{CM}=aE$ can produce heavier states with $M>E$
and $E=E_{1}+E_{2}$ the sum of the two sources' max Fourier  energies. 
As mentioned above, the price to pay for such 
a process is an exponential suppression on 
the cross-section as $a^{-2n}$ for every superoscillating fields participating to the scattering. 
It is worth to remark that superoscillating scatterings would be of interest only for situations where $E T\leq  \sqrt{n}$,
with $T$ the characteristic time of superoscillation propagation considered:
outside this range, superoscillations are totally suppressed. 
We will return on it later in subsection III.D. 
An ordinary resonance from the scattering 
of two superoscillating states with same degree of superoscillation as $a$ (for simplicity)
corresponds to a cross-section 
\begin{equation}
	\label{RE}
	\sigma\simeq \frac{1}{k^{2}}\frac{\Gamma_{i}\Gamma_{f}}{(aE_{CM}-M)^{2}+\Gamma^{2}/2}
\end{equation}
where $E_{CM}$ is the Center of Mass energy, 
$M$ is the resonance mass, $\Gamma, \Gamma_{i},\Gamma_{f}$ are the total and partial decay rates respectively ($\Gamma_{i,f}$
the resonance to initial/final states respectively and $\Gamma=\sum_{j=all}\Gamma_{j}$)
and with $a^{-2n}$ factors to be included for any superoscillating in-states. 

\begin{figure}
	\begin{centering}
		\includegraphics[scale=0.5]{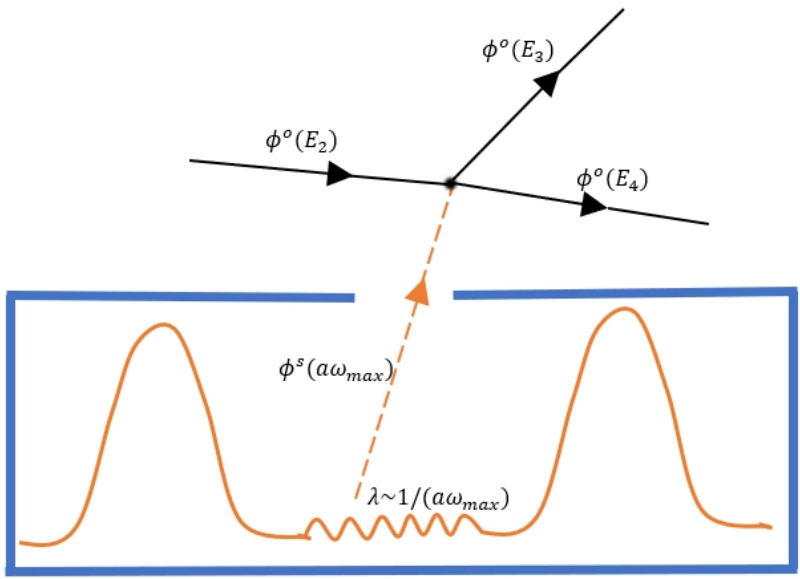}
		\par\end{centering}
	\caption{We show a conceptual example of a superoscillating scalar particle, emitted from 
		a black box source with a max bandwidth frequency $\omega_{max}$. The superoscillating particle elastically scatters with an ordinary particle producing two
		ordinary scalar fields. The superoscillation has a frequency that is $a$-times larger than the bandwidth maximal frequency. }
	\label{fig-Osc-Sto}
\end{figure}

We can consider other exercises for superoscillations and scatterings. 
It is instructive to think about  superoscillations in non-relativistic quantum mechanics scatterings
with the formalism of wave functions. 
As it is known, in non-relativistic quantum mechanics, 
one treats the scattering process 
as a wave superposition of an incident plane wave 
and an outgoing spherical wave as
$$\psi=(2\pi)^{-3/2}(e^{ikz}+f({\bf k'},{\bf k})e^{ikr}/r)$$
where ${\bf k},{\bf k'}$ are in- and out- going wave vectors respectively. 
As it is well known, the differential cross-section is related 
to the amplitude $f$ as $d\sigma/d\Omega=|f|^{2}$. 
In case of an incident superoscillating wave as in Eq.\ref{super}, 
in the limit of $n\rightarrow \infty$, the plane and spherical waves are replaced by superwaves $\sim e^{iakz}$
and $e^{iakr}/r$ respectively. 
In First Order Approximation as well as in $n\rightarrow \infty$ limit, 
the scattering amplitude has the form 
\begin{equation}
	\label{fone}
	f^{(1)}({\bf k},{\bf k'})=-N \int d^{3}x' e^{ia({\bf k}-{\bf k'})\cdot {\bf x}}\, V({\bf x})\, , 
\end{equation}
where superoscillation footprint appears as the ${\it a}$ parameter. $N$ is the prefactor normalization as understood. 
For the energy conservation, $|{\bf k}|=|{\bf k'}|$ and 
$|{\bf k}-{\bf k'}|\equiv q=2k\sin \theta/2$.
After performing integral Eq.\ref{fone},
in case of a Yukawa-like potential 
$V(r)=V_{0}e^{-\mu r}/\mu r$, 
we obtain 
\begin{equation}
	\label{ob}
	f^{(1)}(\theta)=-J \frac{1}{2a^{2}k^{2}(1-\cos \theta)+\mu^{2}}\, . 
\end{equation}
with $J$ understood combination of constants
and 
\begin{equation}
	\label{dsigmad}
	\frac{d\sigma}{d\Omega}=J^{2}\frac{1}{[2a^{2}k^{2}(1-\cos \theta)+\mu^{2}]^{2}}\, . 
\end{equation}
multiplied to an extra $a^{-2n}$ suppression factor. 
Similar energy dependence with respect to the mass pole can be also obtained from Feynman's rules to scattering amplitudes. 
This result indicates that in principle a superoscillating regime scattering can reach interaction mediator mass $\mu$ with 
superoscillating wave sources at max Fourier energies below $\mu$. 
Let us consider another similar example:  an elastic scattering $\phi_{s}(ap_{1})\phi_{s}(ap_{2})\rightarrow \phi_{s}(ap_{3})\phi_{s}(ap_{4})$
with masses $\mu^{2}\phi_{s}^{2}/2$ and interaction vertex $\kappa \phi_{s}^{2}\phi_o$.
In this case, using Feynman's rules, the scattering amplitude through the exchange of a mediator field reads as 
$\mathcal{A}=a^{-2n}(-i\kappa^2)/(q^2+\mu^2)\delta(ap_{1}+ap_{2}+ap_{3}+ap_{4})$
which corresponds to $E_{1}+E_{2}=\mu/a$.  

In Section III.C, we will also discuss the difficulties to obtain scatterings involving superoscillating particles 
in high energy colliders, in relation with the evanescence bound.

\subsection{Laser interferometers and Quantum Gravity}

As pointed out in Ref.\cite{AC}, 
it is possible to probe several Quantum Gravity (QG) models 
from gravitational-wave interferometers.

In particular, one can test QG models with a uncertainty 
scaling as 
\begin{equation}
	\label{sigma}
	\Sigma \sim l_{Pl}^{\alpha}(c\tau)^{1-\alpha}\, ,
\end{equation}
where $\alpha$ is model dependent. 
The standard case considered in QG corresponds to $\alpha=1$, i.e. $\Sigma \sim l_{Pl}$. 
Nevertheless, if QG effects were cumulative,
then $\alpha$ would be different than one. 
This is thought as a possible effect of decoherence from space-time foam. 
The idea is that such an uncertainty would manifest itself 
as a displacement noise in laser interferometers. 

The uncertainty $\Sigma$ is 
\begin{equation}
	\label{sug}
	\Sigma^{2}=\int_{1/\tau}^{f_{max}}|S(f)|^{2}df
\end{equation}
where 
$S(f)$ is the spectral density, $f$ is the frequency and 
$\tau$ is the observation time.
Therefore, Eq.\ref{sigma} corresponds to 
\begin{equation}
	\label{Sfff}
	S(f)\sim c^{1-\alpha}l_{Pl}^{\alpha}f^{\alpha-3/2}\, . 
\end{equation}

Let us now imagine that lasers in the interferometer 
are in a superoscillating states rather than 
ordinary ones. Indeed, we know that superoscillating optical systems 
were already studied in different situations for condensed matter
and material physics, inspiring the possibility to relise the same with lasers. 
In case of superoscillations, in the large-n limit, 
the corresponding effective spectrum would be 
\begin{equation}
	\label{Sfff}
	S(f)\sim a^{3/2-\alpha}c^{1-\alpha}l_{Pl}^{\alpha}f^{\alpha-3/2}\, . 
\end{equation}
Such a spectrum can be formally obtained 
replacing $f_{max}$ in Eq.\ref{sug}
with $f_{max}/a$.

aLIGO has a sensitivity for displacement noise level 
of $10^{-20}\, {\it mHz}^{-1/2}$ around $100\, {\it Hz}$.
In case of the three main branchmark models for quantum gravity, $\alpha=1,2/3,1/2$, 
this corresponds to $l_{Pl}<10^{-17}, 10^{-31},10^{-43}\, {\rm cm}$ respectively. 
In case of an ideal aLIGO concept with superoscillating lasers, 
one can estimate that the bounds would be optimized as 
\begin{equation}
	l_{P l}  \simeq 10^{-\frac{20}{\alpha}} a^{\frac{\alpha-3 / 2}{\alpha}}  c^{\frac{\alpha-1}{\alpha}} f^{\frac{3 / 2-\alpha}{\alpha} }\Big(\frac{m}{\sqrt{Hz}}\Big)^{\frac{1}{\alpha}}, \,\,\,\,\, {\it at}\, f=100Hz\, ,
\end{equation}
which corresponds to 
\begin{equation}
	\label{OP}
	l_{Pl}<a^{-1/2}\, 10^{-17}, \, a^{-5/4}10^{-31},\, a^{-2}10^{-43}\, {\rm cm}\, ,
\end{equation}
for $\alpha=1,2/3,1/2$ respectively.

The last case, the so dubbed random walk model, is already excluded 
for lengths below the Planck scale, while superoscillations 
could increase the sensitivity  for probing the first two QG models. 
Nevertheless, the suppression factor $a^{-n}$ has to be included 
in front of the density spectrum. 
Moreover, our superoscillating interferometers would be efficient 
if lasers on a $\omega L/c<<n$ regime, where $L$ is the laser path distance.
In next sections, we will discuss 
evanescence bounds of superoscillations in various cases, including superoscillating interferometers. 

\vspace{0.2cm}

In order to realize an ideal super-interferometers,
we should not only have lasers in superoscillating 
but also in squeezed states as ones in aLIGO experiment \cite{LIGOsq} (see Ref.\cite{Walls} for a comprehensive introduction to squeezed states). 
The possibility of having a squeezed superoscillating state is theoretically possible 
but never proposed before in previous works. Clearly, the existence of such states 
would have an impact on laser physics which is broader than applications in GW interferometers. 
Here, we propose an example of a superoscillating squeezed electric field 
in the limit of $n\rightarrow \infty$. 
Let us start with a superoscillating electric field in asymptotic limit: 
\begin{equation}
	\label{ddooaa}
	\hat{E}_{s}(t)\rightarrow a\frac{\lambda}{2}(\hat{X}_{1}\cos a \omega t+\hat{X}_{2}\sin a\omega t)
\end{equation}
with a maximal superoscillation frequency $\omega_{s}=a\omega$, 
$a$-factor in front in consistency with QED formulation given in Section II-C,
$\hat{X}_{1,2}$ are operators identifying the amplitudes of the quadrature phases of 
the electric field. Here, $\lambda$ is a constant  which can include the function of space-coordinates. 
For ordinary squeezed states, the uncertainty principle
implies that 
the two quadrant operators 
satisfy the 
quantization condition 
\begin{equation}
	\label{QCQ}
	[\hat{X}_{1},\hat{X}_{2}]=\frac{i}{2}\rightarrow \Delta X_{1}\Delta X_{2}\geq \frac{1}{4}\, ,
\end{equation}
where $\Delta X_{i=1,2}=\sqrt{\Delta V_{i=1,2}}$ and $V_{i,2}(\hat{}X_{i=1,2})$ are the variances
of the $\hat{X}_{i}$-operators.
A squeezed state can be obtained in case the two variances 
are unequal in each quadrants and in particular
\begin{equation}
	\label{varr}
	V(X_{i=1,2})<1/4,\,\,\, i=1\, {\it OR}\, 2\, . 
\end{equation}
Eq.\ref{QCQ} and Eq.\ref{varr} imply that 
one can reduce the variance on one quadrant 
with a cost of increasing the uncertainty 
on the other conjugate one. 
Nevertheless, in case of superoscillations, 
the Heisenberg principle can be eluded 
(with an exponential suppression price cost of the signal intensity $a^{-n}$) 
as 
\begin{equation}
	\label{X1X2}
	V_{1}V_{2}\sim \frac{1}{16a^{4}}
\end{equation}
compatible with equation Eq.\ref{ddooaa} with the a-factor in front of it. 
Eq.\ref{X1X2} can correspond to a super-squeezed state with 
\begin{equation}
	\label{supers}
	V_{1}<\frac{1}{4a^2}
\end{equation}
with a large compensating uncertainty on
$V_{2}$ direction.

Thus we can define a super-squeezed state 
as 
\begin{equation}
	\label{Szeta}
	S(\zeta.a)=\frac{1}{2a}{\rm Exp}\Big(\frac{1}{2}\zeta^{*}(\hat{A})^{2}-\frac{1}{2}\zeta (A^{\dagger})^2 \Big)
\end{equation}
that can be applied on a vacuum or on coherent state as
\begin{equation}
	\label{applied}
	S(\zeta,a)|\psi\rangle =|\psi_{\zeta,a}\rangle\, , 
\end{equation}
where 
\begin{equation}
	\label{AX1X2}
	\hat{A}=\hat{X}_{1}+ i\hat{X}_{2},\,\,\, [\hat{A},\hat{A}^{\dagger}]=1
\end{equation}
are annihilation/creation operators
and 
$\zeta=re^{i\theta}$ is the squeezing parameter 
with $r,\theta$ the squeezing radius and phase respectively. 
Let us note in the limit of $a\rightarrow 1$ the ordinary squeezing operator is re-obtained. 

Defining the operators 
$\hat{Y}_{1,2}=(\hat{X}_{1}+i\hat{X}_{2})e^{-i\theta/2}$,
the variances in squeezed state $|\psi_{\zeta,a}\rangle$
are 
\begin{equation}
	\label{V1V}
	V(Y_{1})=\frac{1}{4a^{2}}e^{-2r}\, ,
\end{equation}
\begin{equation}
	\label{V1V2}
	V(Y_{2})=\frac{1}{4a^{2}}e^{2r}\, ,
\end{equation}
where $2\Delta Y_{1,2}$ 
are the lengths of minor and major axes in 
the uncertainty ellipse. 
Compared to the ordinary squeezed stated, 
the ellipse is reduce of $a^{2}$ factor on both characteristic axes. 

Let us also note that the previous arguments can be easily generalized in case of finite $n$
rewriting Eq.\ref{ddooaa} as follows: 
\begin{equation}
	\label{FREW}
	\hat{E}_{s}(t)=\lambda\sum_{k=0}^{n}c_{k}(n,a)(1-2k/n)[\hat{A}_{k}e^{-i(1-2k/n)\omega t}+\hat{A}_{k}^{\dagger}e^{i(1-2k/n)\omega t}]\, , 
\end{equation}
where now the creation/annihilation operators depend on the $k$-mode. 

An interesting topic of discussion is the following.
As it is known, to squeeze the states is useful as a protection from quantum noise 
which affects coherent states such as lasers in interferometers. 
On the other hand, superoscillations are particularly fragile with respect to external background 
disturbances, including quantum noise. 
Therefore, we conjecture the possibility that squeezing the states can 
render superoscillations more resistant from external backgrounds in quadrant. 

The quantum engineering of superoscillating squeezed states appears to be extremely challenging: it may involve complex superpositions of quantum states or the application of non-linear laser processes that could introduce the desired squeezing and superoscillating characteristics. 
We suggest that an operative way 
to generate a superoscillating squeezed state
is through an interference of N 
identical laser sources, similar to the multi-antenna concept (see the footnote [64]), with a resulting signal passing through a squeezing-state system of crystals. 
This is certainly an experimental frontier which feasibility should be carefully explored in future. 

\subsection{Interaction potentials and quantumness of gravity}

Let us consider the gravitational interaction 
of two scattering particles: 
the amplitude is proportional to 
the gravitational coupling 
$\alpha_{G}(E_{CM})\sim G_{N}E_{CM}^{2}$
where $E_{CM}\sim \sqrt{s}$ is the Center of Mass energy,
$s\equiv (p_{1}+p_{2})^{2}$ is the Mandelstam variable 
and $G_{N}$ is the Newton constant. 
In case of scatterings of superoscillating particles,
in limit of $n\rightarrow \infty$,
the Gravitational coupling is re-scaled 
as $\alpha_{G}\rightarrow a^{2}\alpha_{G}\sim G_{N}(aE_{CM})^2$. 
Such a quadratic deformation remains the same in the non-relativistic limit 
when $E_{CM}\sim m$ with $m$ the mass of the two colliding particles assumed to be the same for simplicity. 
Therefore, in non-relativistic regime, the newtonian potential is re-scaled
as 
\begin{equation}
	\label{rescaled}
	V=\frac{G_{N}m^{2}}{r}\rightarrow a^{2}V=\frac{G_{N}a^{2}m^{2}}{r}\, , 
\end{equation}
corresponding to 
\begin{equation}
	\label{ressss}
	G_{N}\rightarrow G_{N}a^{2}\, . 
\end{equation}

Let us show it with the concrete example of a  $\phi\varphi\rightarrow \phi\varphi$ gravitational elastic tree-level scattering, where $\phi$ and $\varphi$ are two different scalar fields.
Let us consider the graviton-scalar interaction in Minkowki's background as 
$\mathcal{L}_{int}=\frac{\kappa}{2}h_{\mu\nu}T^{\mu\nu}$
with $\kappa=\sqrt{16\pi G_{N}}$. 
The corresponding vertex,
in Feynman's rules, reads as
\begin{equation}
	\label{read1}
	V^{\mu\nu}(k,k')=-\frac{\kappa}{2}(k^{\mu}k'^{\nu}+k'^{\mu}k^{\nu}-\eta^{\mu\nu}(k\cdot k'-m^{2}))\, , 
\end{equation}
where $k,k'$ are momenta of incoming and outcoming scalar field respectively. 
The graviton propagator, in harmonic gauge, is 
\begin{equation}
	\label{harmonicgauge}
	D_{\mu\nu,\rho\sigma}(q)=-\frac{i}{q^{2}+i\epsilon}(\eta_{\mu\rho} \eta_{\nu\sigma} +\eta_{\mu\sigma} \eta_{\nu\rho} - \eta_{\mu\nu} \eta_{\rho\sigma})\, .
\end{equation}
Therefore, the $\phi(ap_{1})\varphi(ap_{2})\rightarrow \phi(ap_{3})\varphi(ap_{4})$ scattering amplitude 
for superoscillating scalar fields 
is 
\begin{equation}
	\label{MMaas}
	\mathcal{M}\simeq V^{\mu\nu}(ap_{1},ap_{3})D_{\mu\nu,\rho\sigma}(q)V^{\rho\sigma}(ap_{2},ap_{4})\, ,  
\end{equation}
where $q=a(p_{1}-p_{3})=a(p_{2}-p_{4})$
is the momentum transferred by the graviton.

In case the scalars $\phi,\varphi$ are distinguishable particles,
only the t-channel contributes to the process
and the scattering amplitude is 
\begin{equation}
	\label{tMM}
	\mathcal{M}\sim G_{N}\frac{a^{4}su}{a^{2}t}=G_{N}a^{2}\Big(\frac{s^{2}}{t}+s\Big)
\end{equation}
with $s=(p_{1}+p_{2})^{2}$, $t=-(p_{1}+p_{4})^2$, 
$u=-(p_{1}+p_{3})^2$ are the Mandelstam variables. 
This amplitude can also be computed using 
the helicity spinor formalism \cite{Bianchi:2008pu}:
\begin{equation}
	\label{shf}
	\mathcal{M}\sim \kappa^2\frac{\mathcal{M}^{MHV}_{4}(1^{-},2^{-},3^{+},4^{+})}{\langle 12\rangle^{8}}\langle 12 \rangle^{2} \langle 13 \rangle^{2} \langle 24 \rangle^{2} \langle 34 \rangle^{2}=s\frac{\langle 13\rangle \langle 24\rangle}{\langle 14 \rangle \langle 32\rangle}\, ,
\end{equation}
where $\mathcal{M}_{4}^{MHV}$ is the well known MHV amplitude of $\mathcal{N}=4$ SYM. 
Eq.\ref{tMM} can be found from Eq.\ref{shf}
multiplying numerator and denominator for 
$[14]$ and using momentum conservation so that 
$\langle 24\rangle [14]=-\langle 23\rangle [13]$,
with $\langle ij\rangle^{2}=a^{2}s_{ij}$
and $s_{12}=s,s_{14}=t,s_{13}=u$.

In non-relativistic limit, the pole 
$a^{2}/t$ corresponds to a Fourier transform in space scaling as $a^{2}/r$: 
\begin{equation}
	\label{FT}
	G_{N}a^{2} m_{\phi}m_{\varphi}\int d^{3}{\bf q}\frac{e^{i{\bf q}\cdot{\bf x}}}{|{\bf q}|^{2}}\sim \frac{G_{N}a^{2}m_{\phi}m_{\varphi}}{r}\, , 
\end{equation}
where $t=|{\bf q}|^{2}$ and $m_{\phi,\varphi}$ are the two scalar field masses.
It is straightforward to check that such a result 
is also obtained in case of indistinguishable 
scalar particles in non-relativistic limit.

Fixing the inter-distance $d$ of two particles 
as a constant, the gravitational potential can be
treated as a constant potential $V_{0}=G_{N}m_{1}m_{2}/d$. 
Let us consider a superoscillating ket state $|\alpha, t_{0};t\rangle$ corresponding to an interaction potential $V(x)$
and a superoscillating $|\alpha,t_{0},t\rangle'$ with a potential 
\begin{equation}
	\label{VVt}
	V'(x)=V(x)+V_{0}\, ,
\end{equation}
where $V_{0}$ is the gravitational potential 
at fixed interdistance $r=d$. 
In case of superoscillations, two states are related each others through a phase transformation: 
\begin{equation}
	\label{VVA}
	|\alpha,t_{0};t\rangle'_{s}=e^{-ia^{2}V_{0}(t-t_{0})/a\hbar}|\alpha,t_{0};t\rangle_{s}\, .
\end{equation}
Here, the $a^{2}$-term is from the potential rescaling in Eq.\ref{FT} and $a^{-1}$
is from $\Delta t=\Delta r/v$
with $v=dE/dp=ap/m$, i.e. 
in non-relativistic limit the propagation 
of super-particle is $a$-times faster than ordinary ones.

As it is well known, in quantum interference phenomena, the phase difference acquires a physical importance.
In case of superoscillations, 
the phase difference induced by gravity reads
as the same of ordinary oscillations with the $a$-rescaling:
\begin{equation}
	\label{pd}
	\phi_{1}-\phi_{2}=a\hbar^{-1}\int_{t_{i}}^{t_{f}}dt[V_{2}(t)-V_{1}(t)]=a\hbar^{-1}V_{0}\Delta t\, ,
\end{equation}
where $V_{0}$ is the gravitational interaction at fixed interdistance, $\Delta t=t_{f}-t_{i}$ and 
interference terms as $\sin(\phi_{1}-\phi_{2})$ and  $\cos(\phi_{1}-\phi_{2})$.

Immediate implications of our results are for tests of quantumness of 
the gravitational field. 
In Ref.\cite{Marletto:2017kzi}, 
the authors proposed a Gedanken experiment with two equal masses $m$ 
which individually undergo to Mach-Zehnder-type interference, 
interacting only through gravity. 
Other interesting proposals on these directions were considered in Refs. \cite{Bose:2017nin,vandeKamp:2020rqh}. 

The main idea is that if the two bodies are entangled 
by gravity, then gravitational interactions will be quantum. 
In the specific set-up described in Ref.\cite{Marletto:2017kzi}
but in case of superoscillating wave functions, we obtain 
modified probabilities to emerge from path ${\bf 0}$ of ${\bf 1}$ for the particles 
as follows: 
\begin{equation}
	\label{pro}
	p_{0}=\frac{1}{2}\Big(\cos^{2}a\phi_{1}^{(o)}/2+\cos^{2}a\Delta \phi^{(o)}/2\Big)\, , 
\end{equation}
\begin{equation}
	\label{prod}
	p_{1}=\frac{1}{2}\Big(\sin^{2}a\phi_{1}^{(o)}/2+\sin^{2}a\Delta \phi^{(o)}/2\Big)\, , 
\end{equation}
where $\phi_{1,2}^{(o)}$ are the relative phases from gravitational potential respectively 
at different interferometer arm distances $d_{1,2}$ while 
$\Delta \phi^{(o)}=\phi_{2}^{(o)}-\phi_{1}^{(o)}$. 
The two states are Maximally entangled if $p_{0}=p_{1}=1/2$, corresponding to 
$\phi_{1}^{(o)}=2n\pi/a$ and $\Delta \phi^{(o)}=\pi/a$. 

In case of superoscillations, the gravitationally induced phases are modified as 
\begin{equation}
	\label{phiii}
	\phi_{i}^{(s)}=(G_{N}a m^{2}/\hbar d_{i})\Delta t=a\phi_{i}^{(o)}
\end{equation}
which can be seen as an a-amplification 
of the gravitational coupling. 

In this sense, tests of quantumness of gravity can be more sensitive 
in a superoscillating interferometer set-up. 
However, it is worth to remark that our theoretical analysis is transcending 
the complexity of producing superoscillating states in such interferometers.
It would be interesting to investigate if there is any experimental concrete prospective which may be realised 
on these directions in next future. For the moment, this can be thought as a {\it gedanken experiment}
for testing conceptual implications of superoscillations in quantumness of gravity.
Moreover, approximated scattering amplitudes in Eq.\ref{tMM} considered above
have extra $a^{-n}$ suppression factors 
which will be relatively close to $1/a$
if saturating the 
{\it Evanescence
	bound}, discussed in next section.

\subsection{Evanescence bound}
Superoscillating modes can survive in a finite time $T$ within the 
range bound 

\begin{equation}
	\label{bound}
	B=|\omega T|< \sqrt{n}\, . 
\end{equation}

For $B>>  \sqrt{n}$, superoscillations 
are exponentially suppressed as $a^{-B^2}$. 
This explains why superoscillation is easily destructed in several observation channels. 
On the other hand, in principle, an experimental apparatus where the frequency is 
set by its size, 
as $B\leq 1$ ($\omega \leq  T^{-1}$),  can efficiently sustain superoscillating modes. 
For example, a system with $B\sim 1$
can encompass superoscillations with 
$a=2$ and $n=10$ as the one showed in Fig.1,
with a suppression factor of $2^{-10}\simeq 0.001$,
eventually compensated by a higher source intensity
with respect to the corresponding ordinary case. 
Moreover superoscillations are states from nearly complete destructive interference of ordinary modes
and therefore they are extremely delicate to external noise perturbations. 

Let us consider the case of superoscillations in laser interferometers, with particular focus on quantum gravity searches as discussed in section III.B. 
In terrestrial experiments such as aLIGO interferometers \cite{aLIGO1,aLIGO2},
the typical frequency range is 
$f \sim 1\div 100\, Hz$
while the time scale $T$ is related 
to the arm length scale as $T=R/c\sim 10^{-5}\,{\rm s}$. 
Using Eq.\ref{bound}, this corresponds
to $B=(10^{-5}\, s)(1\div 100\, Hz)\sim 10^{-5}\div 10^{-3}$. Thus, for superoscillating lasers, the B-factor is $B<<1$ for all the frequency range of interest. 
This suggests that in principle an aLIGO-like concept with superoscillating laser beams rather than ordinary ones can probe  shorter 
QG length scales in same frequency range of 
$1\div 100\, {\rm Hz}$, with model dependent parametrization as in Eq.\ref{OP}. 
Such an estimation can be taken as "good news" but the technical feasibility of such a superoscillating 
interferometer has to be explored yet. 

Similar analysis can be extended to the third generation of European GW detectors, the Einstein Telescope (ET) \cite{ET1,ET2} and the Cosmic Explorer (CE) \cite{CE1,CE2}.  
Indeed, these detectors, compared to aLIGO, will lose few numerical factors in length inside the evanescence bound $B$ largely overcompensated from the increase of sensitivity, while exploring a similar frequency range. 

On the other hand, in case of space-based experiments such as eLISA \cite{LISA}, 
the arm is 2.5 million kilometres or so, corresponding to the typical time scale of about $10\, {\rm s}$.
Thus $B \sim (10\, s)(10^{-4}\div 10^{-2}\, Hz)\sim 10^{-3}\div 10^{-1} \ll 1$. Therefore, a hypothetical LISA-like experiment equipped with superoscillating lasers would be effective in comparison with the evanescence limit. 

Other proposals for quantum gravity tests are table-top interferometers 
of size $\sim 10\,{\rm m}$
and frequency of $1\,{\rm MHz}\div 1\,{\rm GHz}$ (see for example Refs.\cite{MHZ1,Aggarwal:2020olq} ). 
In this case, the time scale is about $10^{-7}s$, thus $B \sim (10^{-7}\, s)(10^{6}\div 10^{9}\, Hz)\sim 0.1\div 100$. Thus such detectors with superoscillating lasers would work well in MHz frequency region while would become ineffective for higher frequencies around GHz.

Concerning astrophysical sources, 
superoscillating signals that can arrive to observers in proximity of the Earth 
are with $B=\omega R/c$,
where $a\omega$ is the characteristic  frequency of the propagating superoscillating signal in large-n limit
and $R$ is the source distance from detector. 
Thus superoscillating messengers from astrophysical sources are effectively lost. 
For instance, superoscillations of electromagnetic or gravitational waves with $\omega \sim {\rm Hz}$ would decay only after 
one light-sec as faster than $a^{-R/d}$,
where $R$ is the astrophysical source distance and $d\sim (1/\omega)c$.

For high energy scatterings, 
the propagation distance $d$ of a superoscillating particle has to be lower or of the same order of $h/cE$
where $E$ is its characteristic energy.
Such a physical regime seems to be 
disfavoured in any conventional high energy colliders, where particles are accelerated for a relatively long distance before reaching a high kinetic energy. For example in LHC,
the TeV-scale energy
against several kms of particle beam propagation time corresponds to a $B>>1$ of 18-19th digits or so. On the other hand, it may be interesting to explore implications of superoscillating fields, produced as short-living particles or resonances, decaying to other Standard Model (SM) particles. 

An important aspect to consider is how superoscillations may be, in certain circumstances, "lost and regained"
through interactions with ordinary SM particles, transferring energy and information, before the evanescence. 
Interactions of ordinary and superoscillating fields can also generate quantum entanglement 
which allows to transfer quantum  informations of superoscillations at larger distances than the superoscillating bounds. 
Some aspects of quantum entanglement of superoscillations were also discussed in case of gravity-mediation in Section above.  
On the other hand, one can realise a quantum fully entangled state with a wave function that is superoscillating.
Superoscillating entangled states were not exhaustively analysed yet and they deserve  further investigations beyond 
the purposes of this work. 
Another hypothetical possibility,
subtly interplaying with the evanescence, 
would be that 
quantum superoscillating fluctuations in early Universe 
were amplified during inflation.
Just as inflation can stretch regular quantum fluctuations, it could also distend primordial superoscillations. Due to the exponential expansion of space itself, these superoscillating modes would be expanded to macroscopic scales, potentially leaving imprints across the Universe at scales much larger than those of the initial fluctuations \footnote{
	The amplification of quantum superoscillations 
	can be studied from the Mukhanov-Sasaki (MS) equation,
	which governs the evolution of quantum perturbations of a scalar field in expanding Universe background. As it is known, the MS equation describes how quantum fluctuations  evolve during the inflationary period and get stretched to astronomical scales, laying the groundwork for the formation of cosmic structure \cite{Mpaper,Spaper}.
	The MS equation is $v''_{K}+(K^{2}-z''/z)v_{K}=0$
	where $v_{K}=z \mathcal{R}_{K}$, $z=a\dot{\phi}/H$
	with $\phi$ the inflaton field, $a$ the scale factor of the Universe, $H$ the Hubble rate,
	$\mathcal{R}_{K}$ the Fourier transform of the comoving curvature perturbations. $v''$ denotes the second derivative with respect to the conformal
	time $d\eta=dt/a$ with $t$ the cosmological time;
	$K$ represents the wave-number of the perturbation, related to its inverse spatial scale; $z''/z$ encapsulates the effect of the expanding universe on the evolution of the perturbations, effectively acting as a time-varying mass term in MS equation. While the mathematical form of the MS equation is the same for ordinary or superoscillating 
	perturbations, the time evolution of it will depend from the function input. 
	A general solution of the MS equation is 
	difficult due to the background dependence in the $z''/z$ part, and for superoscillations one can expect an even more challenging problem.}.
In this case, their effects 
may be relevant for late Universe cosmology 
even if only surviving at much shorter time
than the Universe age,
and impacting on the Cosmic Microwave Radiation (CMB)
with new characteristic signatures. 
A full investigation of superoscillations in early Universe cosmology is far beyond the purpose of this paper. 

Therefore, the evanescence bounds render superoscillations elusive at long distances for a direct detection
but, in principle, their indirect effects would survive under favorable conditions.

\subsection{Superoscillations of particles}

Let us consider the 
Schr\"odinger equation of two particles 
\begin{equation}
	\label{SHH}
	i\hbar \frac{d}{dt}\Psi=H \Psi\, , 
\end{equation}
where 
\begin{equation}
	H = \begin{pmatrix}
		\epsilon_{11} & \epsilon_{12}  \\
		\epsilon_{12}^{*} & \epsilon_{22}  \\
	\end{pmatrix}\, ,\,\,\, \Psi(t)=\begin{pmatrix}
		\psi_{1}(t) \\
		\psi_{2}(t) \\
	\end{pmatrix}\, . 
\end{equation}
Let us consider an initial state for 
this equation as
\begin{equation}
	\label{COrre}
	\Psi(0)=\begin{pmatrix}
		\Big(\cos \frac{n{\bf p}\cdot {\bf x}}{\hbar}+i\,a\, \sin \frac{n{\bf p}\cdot {\bf x}}{\hbar}\Big)^{n} \\
		0 \\
	\end{pmatrix}\,
\end{equation}
corresponding to $\Psi(0)=(e^{ia{\bf p}{\bf x}},0)^{T}$ in the limit of $n\rightarrow \infty$. 
Eq.\ref{COrre} represents an initial state prepared as superoscillating on one of the two particles. 

The Hamiltonian 
can be diagonalised, with eigenvalues 
\begin{equation}
	\label{eigenL}
	E_{\pm}=\frac{1}{2}\big[\epsilon_{11}+\epsilon_{22}\pm \sqrt{(\epsilon_{11}-\epsilon_{22})^{2}+4|\epsilon_{12}|^{2}}\big]
\end{equation}
and 
eigenkets $|\pm\rangle$ as 
\begin{equation}
	\label{eigen}
	|+\rangle =e^{-i\phi/2}\cos \frac{\theta}{2}|1\rangle +e^{i\phi/2}\sin \frac{\theta}{2}|2\rangle\, ,
\end{equation}
\begin{equation}
	\label{eigen2}
	|-\rangle =-e^{-i\phi/2}\sin \frac{\theta}{2}|1\rangle +e^{i\phi/2}\cos \frac{\theta}{2}|2\rangle\, ,
\end{equation}
where $\theta,\phi$ are the mixing angle and phase respectively. 

The time-evolution of the state $\Psi(t)$
is 
\begin{equation}
	\label{timeevo}
	|\Psi(t)\rangle=e^{iHt/\hbar}|\Psi(0)\rangle=e^{i\phi/2}\Big(\cos \frac{\theta}{2}|+\rangle e^{-\frac{i}{\hbar}E_{+}t} -\sin \frac{\theta}{2}|-\rangle e^{-\frac{i}{\hbar}E_{-}t} \Big)\, .
\end{equation}

The probability for a $1\rightarrow 2$
transition is 
\begin{equation}
	\label{Probn}
	P_{21}(t)=|\langle 2|\Psi(t)\rangle|^{2}=\sin^{2}\theta \sin^{2}\omega t\, , 
\end{equation}
with 
\begin{equation}
	\label{OSCC}
	\sin^{2}\theta=\frac{4|\epsilon_{12}|^{2}}{4|\epsilon_{12}|^{2}+(\epsilon_{11}-\epsilon_{22})^{2}}\, , 
\end{equation}
and 
\begin{equation}
	\label{OMEGAA}
	\omega=\frac{E_{+}-E_{-}}{2\hbar},
\end{equation}
with $E_{+}-E_{-}=\sqrt{4|\epsilon_{12}|^{2}+(\epsilon_{11}-\epsilon_{22})^{2}}$.

In a full relativistic superoscillating regime, 
\begin{equation}
	\label{Epmm}
	\epsilon_{\alpha\alpha}=\sqrt{p^{2}_{\alpha}+m_{\alpha \alpha}^{2}/a^{2}}\, , 
\end{equation}
with $\alpha=1,2$. 
In super-relativistic regime, 
Eq.\ref{Epmm} can be approximated as 
\begin{equation}
	\label{Epmm}
	\epsilon_{\alpha\alpha}\simeq p+\frac{m_{\alpha}^{2}}{2a^{2}E}\, . 
\end{equation}
Inserting this relation inside Eq.\ref{OSCC} and Eq.\ref{OMEGAA}.
we obtain 
\begin{equation}
	\label{SIInn}
	\sin^{2}\theta_{a}=\frac{4|\epsilon_{12}|^{2}}{4|\epsilon_{12}|^{2}+(\Delta m_{\alpha}^{2}/2a^{2}E)^2}\, ,
\end{equation}
and 
\begin{equation}
	\label{maiai}
	\omega_{a}=\sqrt{4|\epsilon_{12}|^{2}+(\Delta m_{\alpha}^{2}/2a^{2}E)^2}\, .
\end{equation}
From Eq.\ref{SIInn}, it is clear that the a-parameter
effectively amplifies the mass mixing parameter compared to the diagonal mass values,
i.e. it effectively increases the mixing angle
with respect to ordinary cases. 
However, the suppression $a^{-n}$ factor for the superoscillating wave function has to be taken into account as 

\begin{equation}
	\label{Probnsupp}
	P_{21}(t)=|\langle 2|\Psi(t)\rangle|^{2}\simeq a^{-2n}\sin^{2}\theta_{a} \sin^{2}\omega_{a} t\, . 
\end{equation}

\subsubsection{Neutrino superoscillations}
Let us consider the case of neutrinos.
As it is known, neutrinos of different flavors 
mix in time due to the fact that flavor eigestates are not the same eigenstates of the mass matrix.
In particular, the states $|\nu_{\alpha}\rangle$
and $|\nu_{i}\rangle$, with 
$\alpha,\beta$ flavors, $i,j$ mass eigenstate indices, are related each others as 
\begin{equation}
	\label{evone}
	|\nu_{\alpha}(t)\rangle=\sum_{i}U_{\alpha i}^{*}|\nu_{i}(t)\rangle\, , 
\end{equation}
and the $\alpha\rightarrow \beta$ transition probability in time is 
\begin{equation}
	\label{PPPPRROb}
	P_{\alpha\beta}=|\langle \nu_{\beta}|\nu_{\alpha}(t)\rangle|^{2}=|\sum_{i}\sum_{j}U^{*}_{\alpha i}U_{\beta j}\langle \nu_{j}|\nu_{i}(t) \rangle|^{2}\, , 
\end{equation}
with $|\nu_{i}\rangle=e^{-iE_{i}t}|\nu_{i}(0)\rangle$
with $E_{i}=\sqrt{p_{i}^{2}+m_{i}^{2}}$
corresponding to energy eigenvalues depending on mass eigenvalues. 
In case of superoscillating flavor states, 
the $m_{i}$ non-trivially scales with $a$-parameter
as eigenvalues of the effective matrix 
with diagonal terms $m_{\alpha\alpha}/a^{2}$.
However, such an effect seems to be not directly observable 
from neutrino oscillation experiments since the relevant 
parameters tested 
are the differences of squared mass matrix eigenstates. 
As an alternative, we consider
mass eigenstates in superoscillations: 
in this case, 
\begin{equation}
	\label{conss2}
	|\nu_{i}\rangle\simeq e^{-iE_{i}(a)t}|\nu_{i}(0)\rangle,\,\,\, E_{i}(a)=\sqrt{p_{i}^{2}+m_{i}^{2}/a^{2}}\, . 
\end{equation}
In the limit of relativistic superoscillating neutrinos, Eq.\ref{conss2} corresponds to
$E_{i}\simeq E\simeq p+\frac{m_{i}^{2}}{2a^{2}E}$ with $p_{i}\simeq p$.
Thus, the two neutrino superoscillation
has a modified transition probability as 
\begin{equation}
	\label{prob2}
	P_{\alpha \rightarrow \beta,\alpha\neq \beta}\simeq a^{-2n}\sin^{2}(2\theta_{a})\sin^{2}\frac{\Delta m^{2}L}{4a^{2}E}\, ,
\end{equation}
where the $a^{-2n}$ is the usual superoscillating suppression factor, $\Delta m^{2}=m_{2}^{2}-m_{1}^{2}$ ($i=1,j=2$ for the first and second mass eigenstate), $E$ is the neutrino energy and $L$ is the propagation distance
from the source to the detector. 
The mixing angle corresponds to the 
$2\times 2$ $U$-matrix for the two neutrinos reading as 
\begin{equation}
	U=\begin{pmatrix}
		\cos \theta_{a} & \sin \theta_{a} \\
		-\sin \theta_{a} & \cos \theta_{a} \\
	\end{pmatrix}\, \, .
\end{equation}
The corresponding neutrino propagation length 
in Eq.\ref{prob2}
is 
\begin{equation}
	\label{Promm}
	L_{0}^{(s)}=\frac{4\pi a^{2}E}{\Delta m^{2}}=a^{2}L_{0}^{(o)}\, ,
\end{equation}
where $(s),(o)$ correspond to super- and ordinary- lengths respectively. 
These results suggest that if mass eigestates 
are in superoscillating states,
then the characteristic oscillation length in Eq.\ref{Promm} will be longer, rather than shorter 
as one may naively expect.
In fact, the squared mass difference is effectively 
reduced of a factor $a^{2}$ within the conversion probability Eq.\ref{prob2}. However, the $a^{-2n}$
factor has to be considered, corresponding to the evanescence bound $B=|\omega T|<\sqrt{n}$. Nevertheless,
here $B$ corresponds to the  argument of Eq.\ref{prob2}, i.e. 
\begin{equation}
	\label{Bneutrino}
	B=\frac{\Delta m^{2}L}{4E}=1.267\frac{\Delta m^{2}}{eV^{2}}\frac{L/E}{meter/MeV}<\sqrt{n}\, .
\end{equation}

For $L<<a^{2}L_{0}^{(o)}$, Eq.\ref{prob2} is suppressed as 
$(L/a^{2}L_{0}^{(o)})^{2}$ not giving any appreciable effects;
for $L>>a^{2}L_{0}^{(o)}$ the probability is average
on any oscillation cycles with $\langle \sin^{2}(L/a^{2}L_{0}^{(o)})\rangle =1/2$. 

Such considerations can be easily generalised in case
of three neutrinos: the transition probability reads as 
\begin{equation}
	\label{Pallaa}
	P_{\alpha\beta}=\delta_{\alpha\beta}-4\sum_{i<j}{\rm Re}[U_{\alpha i}U_{\beta i}^{*}U_{\alpha j}U_{\beta j}]\sin^{2}\frac{X_{ij}}{a^{2}}+2\sum_{i<j}{\rm Im}[U_{\alpha i}U_{\beta i}^{*}U_{\alpha j}U_{\beta j}]\sin 2\frac{X_{ij}}{a^{2}}\, ,
\end{equation}
with 
\begin{equation}
	\label{Xijjj}
	X_{ij}=\frac{(m_{i}^{2}-m_{j}^{2})L}{4E}\simeq 1.267\frac{\Delta m_{ij}^{2}}{{\rm eV}^{2}}\frac{L/E}{meter/MeV}\, . 
\end{equation}

\begin{table}[htbp]
	\centering
		\begin{tabular}{|l|c|c|c|c|}
			\hline
			Experiment & $L$ (m) & $E$ (MeV) & $|\Delta m^2|$ (eV$^2$) & B \\
			\hline
			Solar & $10^{10}$ & 1 & $10^{-10}/a^{2}$ & $B\sim 1$\\
			\hline
			Atmospheric & $10^4 - 10^7$ & $10^2 - 10^5$ & $10^{-1}/a^{2} - 10^{-4}/a^{2} $ & $B\sim 1$\\
			\hline
			Reactor & \begin{tabular}{@{}c@{}} SBL:\, $10^2 - 10^3$ \\ LBL:\, $10^4 - 10^5$ \end{tabular} & 1 & \begin{tabular}{@{}c@{}} $10^{-2}/a^{2} - 10^{-3}/a^{2}$ \\ $10^{-4}/a^{2} - 10^{-5}/a^{2}$ \end{tabular} & $B\sim 1$ \\
			\hline
			Accelerator & \begin{tabular}{@{}c@{}} SBL:\, $10^2$ \\ LBL:\, $10^5 - 10^6$ \end{tabular} & \begin{tabular}{@{}c@{}} $10^3 - 10^4$ \\ $10^3 - 10^4$ \end{tabular} & \begin{tabular}{@{}c@{}} $\geq 0.1/a^{2}$ \\ $10^{-2}/a^{2} - 10^{-3}/a^{2}$ \end{tabular} & $B\sim 1$ \\
			\hline
	\end{tabular}
	\caption{Summary of neutrino experiments and their parameters \cite{PDGNeutrino} thinking about hypothetical superoscillating neutrinos. In all these channels, the B-factor is close to one, easily within the evanescence bound.  }
	\label{tab:neutrino_experiments}
\end{table}

In principle, superoscillations can probe 
neutrino mass differences which are normally not accessible from ordinary ones at given energies and propagation lengths. A summary is shown in TABLE I, 
considering hypothetical superoscillating neutrinos 
from several sources.
Clearly, such estimations 
do not capture the whole complexity of detecting superoscillating neutrinos in experiments. 
For example, it is reasonable that superoscillations from solar, atmosferic and astrophysical neutrinos 
are rarely generated and largely suppressed.
It may be possible to realise a more controllable production of superoscillating neutrinos in laboratories 
such as in Short-Base-Lines (SBL) and Long-Base-Lines (LBL) with particle accelerators, producing superoscillating neutrinos from scattering of accelerated particles on a target. 
In these experiments, the scattering secondaries 
such as pions and kaons decay out producing neutrinos. If initial protons were set to a superoscillating state, then in principle superoscillating neutrinos could be generated. 
However, such possibilities remain speculative and far beyond a concrete technical project.

\subsubsection{Kaon-Antikaon superoscillations}

As a next relevant example, let us consider Kaon-Antikaon Superoscillations.
As it is known, Kaon-Antikaon oscillations serve as a profound example of quantum mechanical phenomena in particle physics, illustrating not only the violation of CP symmetry but also providing insight into the interaction between particles and their antiparticles.
The $K^0 - \bar{K}^0$ transition arises due to the weak interactions that allow the neutral kaon eigenstates to convert into each other, violating strangeness conservation. The $\bar{K}^0$ state is defined as the CP conjugate of the $K^0$ state:
\begin{equation}
	|\bar{K}^0\rangle = CP|K^0\rangle\, .
\end{equation}

The evolution of the system is governed by the non-relativistic Schr\"odinger equation:
\begin{equation}
	i\hbar \frac{\partial}{\partial t} \begin{pmatrix} K^0 \\ \bar{K}^0 \end{pmatrix} = \mathbf{H} \begin{pmatrix} K^0 \\ \bar{K}^0 \end{pmatrix},
\end{equation}
where $\mathbf{H}$ is the effective Hamiltonian matrix, not necessarily hermitian due to weak interactions.
As it is known, 
$K_{0}-\bar{K}_{0}$
are extremely sensitive to possible CPT violations which can be induced 
from non-commutative quantum gravity models such as $\kappa$-Poincar\'e, or from decoherence 
induced by space-time foam. 
If CPT is violated, 
the diagonal terms 
in the Hamiltonian,
including masses 
and decay rates, 
will be different.
The effective Hamiltonian 
for $K_{0}-\bar{K}_{0}$
reads as 
\begin{equation}
	\mathbf{H} = \begin{pmatrix} m_{11} - i\frac{\Gamma_{11}}{2} & m_{12} - i\frac{\Gamma_{12}}{2} \\ m_{12}^* - i\frac{\Gamma_{12}^*}{2} & m_{22} - i\frac{\Gamma_{22}}{2} \end{pmatrix},
\end{equation}
where $m_{11,22}$ and $\Gamma_{11,22}$ are the masses and decay widths of the kaon and antikaon, respectively, and $m_{12}$, $\Gamma_{12}$ account for mixings and decays.
If CPT is preserved,
then $m_{11}=m_{22}$ and $\Gamma_{11}=\Gamma_{22}$.
We can rewrite 
$m_{11,22}=m\pm \frac{1}{2}\delta m$
and 
$\Gamma_{11,22}=\Gamma\pm \frac{1}{2}\delta \Gamma$,
where $\delta m,\delta\Gamma$
represent the deviations from 
the CPT preserving case \cite{EllisCPT}. 
Let us note that in most of 
CPT violating quantum gravity models $\delta m,\delta \Gamma$
are functions of characteristic momenta or the energies 
involved in the process.
Therefore, considering $K_{0}-\bar{K}_{0}$ in superoscillations 
can amplify the sensitivity to CPT violations as a model dependent effect. 
As an example, let us consider the case of $\kappa$-Poincar\'e:
the modified dispersion relations (MDRs) \cite{Amelino-Camelia:2000ikd} 
are 
\begin{equation}
	\label{MDRs}
	c^{4}m^{2}=\frac{\hbar^{2}c^{2}}{\lambda^{2}}\Big(e^{\lambda E/\hbar c}+e^{-\lambda E/\hbar c}-2 \Big)-c^{2}|{\bf p}|^{2}e^{-\lambda E/\hbar c} ,
\end{equation}
where $\lambda=\hbar c/\Lambda$
and $\Lambda$ the new-physics energy scale. It is worth to note
that MDRs have also complementary 
phenomenology in underground experiments 
such as BOREXINO, DAMA, VIP etc in relation to violations
of the Pauli Exclusion Principle \cite{Addazi:2017bbg}
and very high energy gamma rays from Gamma Ray Bursts (see \cite{Addazi:2021xuf} for a complete review). 
In case of superoscillations, 
in large-n limit,
Eq.\ref{MDRs}
is re-scaled as 
\begin{equation}
	\label{MDRs2}
	c^{4}m^{2}\simeq \frac{\hbar^{2}c^{2}}{\lambda^{2}}\Big(e^{a\lambda E/\hbar c}+e^{-a\lambda E/\hbar c} \Big)-c^{2}a^{2}|{\bf p}|^{2}e^{-a\lambda E/\hbar c}\simeq a^{2}(E^{2}-c^2|{\bf p}|^{2})+a^{3}\frac{\lambda c}{2\hbar}E |{\bf p}|^{2}\, .
\end{equation}
Eq.\ref{MDRs2} has two solutions,
the positive energy one corresponds to 
\begin{equation}
	\label{TwoSOl}
	E\simeq  \sqrt{m^{2}c^{4}/a^{2}+c^2|{\bf p}|^{2}}-\frac{a\lambda c}{\hbar}|{\bf p}|^{2}\, ,
\end{equation}
(neglecting the $\lambda^2$-term). 
Therefore, a momentum dependent 
CPTV mass splitting 
is generated by Quantum Gravity Effects
as follows:
\begin{equation}
	\label{splitting1}
	\frac{|\delta m|}{E}\sim \frac{a\lambda c|{\bf p}|^{2}}{\hbar \sqrt{c^{2}|{\bf p}|^{2}+m^{2}c^{4}/a^{2}}}\, ,
\end{equation}
and $\delta \Gamma/\Gamma \sim 4\delta m/E$. 
As a consequence of the 
fact that 
$\Gamma/m<<1$,
the main test 
for CPT is from the mass splitting
in Eq.\ref{splitting1}.
Moreover, the 
CPTV mass splitting 
competes 
with the 
mass difference 
of the 
mass eigenstates 
$m_{K_{L}}-m_{K_{S}}\sim 3\times 10^{-15}\, {\rm GeV}$. 
Infact, as it is known, the eigenstates $K_L$ (long-lived) and $K_S$ (short-lived) of the Hamiltonian are mixtures of $K^0$ and $\bar{K}^0$, showing different lifetimes and mass eigenvalues.

Let us note that Eq.\ref{splitting1}
predicts a change of a-dependence in 
ultra-relativistic and non-relativistic regimes 
from $a$ to $a^{2}$
while in between 
a function that 
at fixed kinetic/mass $(K/m)$
scales ad $f(a)=a^{2} (K/m)/\sqrt{a^{2}(K/m)+1}$. 
The price to pay for the intensity 
is at the best $a^{-n}$ for $n\simeq O(B)\sim O(1)$, which may be recovered with larger statistics.
In Table II we summarise 
how superoscillation
could improve collider constraints 
on CPTV if it was relisable 
with the same experimental concepts. 
Electron-Positron colliders for Meson factories ${\rm DA\Phi NE}$ provide
a powerful experimental test for CPT 
\cite{DAFNE,DiDomenico}. For instance, KLOE experiment
can test Kaon-Antikaon systems 
through the $\phi$-resonance \cite{DAFNE}.
CPLEAR experiment tests kaon-physics 
from proton-antiproton scatterings 
as $p\bar{p}\rightarrow K^{\pm}\pi^{\mp}(\bar{K}^{0},K^{0})$ \cite{CPLEAR}.
NA31 at CERN \cite{NA} and E731 at Fermilab \cite{E7}
studied kaon decays into pions 
searching for CP and CPT violations,
with kaons produced from a proton beam 
scattered on a target.

Let us remark that here we do not pretend to use the past and current experiments for superoscillations:
for the moment our estimations have to be cautiously taken as exercises which may inspire future collider concepts.
From our theoretical analysis, it emerges that
superoscillations in Kaon-Antikaon 
can amplify CPT violations 
from Quantum Gravity,
with hypothetical applications 
in  new experiments. 

\begin{table}[htbp]
	\centering
		\begin{tabular}{|l|c|c|c|}
			\hline
			Experiment & $E$ (GeV) & $|\delta m|$ (GeV) & $\lambda$(meters) \\
			\hline
			KLOE & $\sim 0.1$& $3\times 10^{-18}$ & $6\times 10^{-32}/f(a) $\\
			\hline 
			CPLEAR & $\sim 0.6$ & $7.5\times 10^{-18}$ & $1.2\times 10^{-32}/f(a)$\\
			\hline
			NA31 \& E731 &$ \sim 10$& $5\times 10^{-18}$ &$ 10^{-35}/a$\\
			\hline
	\end{tabular}
	\caption{Summary of bounds CPTV in $K_{0}-\bar{K}_{0}$ imaging the possibility of similar experimental concepts realizing superoscillations. 
		In particular, we compare the average energy $E$ of the produced Kaons, the mass splitting $\delta m$ and the limit on the $\lambda$-parameter. }
	\label{tab:kaon_experiments}
\end{table}

\subsection{Tunneling from barrier potentials}

Let us consider superoscillating wave functions 
in presence of a constant potential 
\begin{equation}
	\label{constant}
	V(x)=V_{0}\,\,\,\, x_{1}\leq x\leq x_{2}\, ,\,\,\, V(x)=0\,\,\, {\it elsewhere}\, . 
\end{equation}
The corresponding action for this potential is 
\begin{equation}
	\label{action}
	\gamma=\int_{x_{1}}^{x_{2}}dx\sqrt{\frac{2m(V(x)-E)}{\hbar^{2}}}=\sqrt{\frac{2m(V_{0}-E)}{\hbar^{2}}}\, . 
\end{equation}

This result can be applied for superoscillating wave functions. 
In case of fixed $n$ we should consider all the possible 
$p_{k}=(1-2k/n)p,E=p_{k}^2/2m$ and several $\gamma_{k}$-contributions to the action.
The overall tunneling behavior of the superoscillating wave would be a complex interference pattern of the tunneling probability amplitudes of k-components.  
But in the large-n limit, the expression Eq.\ref{action} is approximately modified as 
\begin{equation}
	\label{actionV2}
	\gamma_{s}\sim \sqrt{\frac{2m(V_{0}-a^{2}p^{2}/2m)}{\hbar^{2}}}\, , 
\end{equation}
with the superoscillating wave function as
approximately free-propagating inside the box, with a momentum $ap$. 
We also consider the barrier with a size
much below the limit from the evanescence bound. 
Thus, $\gamma_{s}<\gamma_{o}$ since
\begin{equation}
	\label{min}
	V_{0}-2a^{2}E<V_{0}-2E\, ,
\end{equation}
with $E=p^2/2m$.
Therefore the tunneling transition probability rate for superoscillating modes,
compared to the normal one, reads as 
\begin{equation}
	\label{higher}
	\frac{\Gamma_{s}}{\Gamma_{o}}\sim e^{-\gamma_{s}+\gamma_{o}-2n\log a}\, .
\end{equation}
where $2n\log a$ is from the typical superoscillation damping factor $a^{-2n}$. 
Such a ratio of transition rates is exponentially sensitive to the $a$ parameter, as evident from Eq.\ref{higher}.

As it is known, tunneling processes correspond to instantons.
The fact that the tunneling probabilities are different for superoscillating waves suggests a non-trivial modification of instanton solutions in 
case of finite-$n$. In the limit of $n\rightarrow \infty$, instantons are expected to be rescaled just by an $a$-factor.
Here, we will show that, in WKB quantum mechanics, instantons related to superoscillations 
have the same form and profile that ordinary ones. 
We put forward a specific case in QFT: let us consider a
$0+1$ scalar field theory with a double field potential.  
In order to find an instanton solution, we first consider the Euclidean action, which involves a Wick rotation $t\rightarrow -i\tau$.

In the limit of almost free propagating particle 
inside the potential, the Euclidean action reads as 
\begin{equation}
	\label{SEE}
	S_{E} \sim \int d\tau \Big[a^{2}\frac{1}{2}\Big( \frac{d\phi}{d\tau}\Big)^{2}+\frac{\lambda}{4}(\phi^{2}-v^{2})^{2}\Big]\, , 
\end{equation}
valid in regime where  $\lambda v^{2}\phi^{2}>> \lambda v \phi^{3},\lambda \phi^{4}$ (dominating mass term on other self-interactions).

The equation of motion derived from $\delta S_{E}=0$
corresponds to 
\begin{equation}
	\label{relaa}
	\frac{d^{2}\phi}{d\tau^{2}}=-a^{-2}\frac{dV}{d\phi}=-a^{-2}\lambda(\phi^{2}-v^{2})\, , 
\end{equation}
and it 
leads to the classical paths in the inverted potential scenario,
in case of superoscillations.
For an instanton solution, which describes tunneling from one well to another, we look for a solution that connects the two minima $\pm v$.
Setting the mass as $m=1$, instantons have profile 
\begin{equation}
	\phi(\tau)\rightarrow v\tanh(\omega_{a} \tau),\,\,\, \omega_{a}=\sqrt{\lambda v^{2}/2a^{2}}\, .
\end{equation}
This means that the instanton form function is still the same 
of the ordinary case but the frequency $\omega$ is re-scaled 
as a factor $a^{-1}$.
From Eq.\ref{relaa}, $a$ is lowering the characteristic energy $\omega$
rendering the instanton variation 
around $\tau=0$ smoother than the ordinary one. This is equivalent to an effective lowering of the separation barrier from decreasing the ratio of $m$ over $\lambda$ parameters. 
On the other hand, for $k=0,..,n$ with finite $n$,
a sum on all over k-instantons is expected. 
The re-interpretation of WKB in case of superoscillations from the prospective of instantons would suggest that a possible extension to more complicated cases in QFT, such as
in electroweak and QCD theories, could be envisaged, beyond our preliminary simple analysis. Moreover, the fact that instantons are in correspondence with solitons in higher dimensions 
(see for example Ref.\cite{Addazi:2016yre}) suggests the existence of a class of a-deformed solitons including monopoles living within a lifetime limit fixed by the evanescence bound. 

Superoscillations and tunneling were discussed in relations 
with superluminality (see Refs. \cite{Superluminal1,Superluminal2,Superluminal3,Superluminal4}).
Indeed, superoscillation tunnelings in particular potentials,
such as a large number of $\delta$-Dirac, 
can be connected to studies of superluminal group velocities. 
This does not violate causality since, in these cases, the velocity group $v_{g}=d\omega/dk$ is 
not strictly identified with the signal velocity and it cannot transport any 
classical informations in useful way \cite{Superluminal1,Superluminal2,Superluminal3,Superluminal4}. 
Such considerations on superluminality can in principle be extended
from non-relativistic quantum mechanics to the case of relativistic quantum field theories.

Eq.\ref{higher} inspires to consider the case
of speculative superoscillation modes tunneling out from a Black Hole.
This is highly motivated from thinking about a BH as an ideal Black Body, in turn motivating the existence of superoscillations as in Ref.\cite{Super1}. 
In next section, we provide an argument 
on how superoscillating radiation could be emitted from the BH horizon as a leading order effect comparable to Hawking's emission.

\subsubsection{Superoscillating tunneling outside Black Holes and Hawking's radiation}
One heuristic way to derive the Hawking's temperature \cite{Hawking1,Hawking2} is based on  the uncertainty principle (see for example Ref.\cite{Pino}).
According to Heisenberg's uncertainty principle, the appearance-disappearance time
of a pair of virtual particles corresponds to 
\begin{equation}
	\label{unc}
	\Delta t\sim \hbar/\Delta E=\hbar/E\, , 
\end{equation}
where $\Delta E= E$ is the energy of the particles.
This implies that the maximal travel distance of 
a virtual particle before disappearing 
is just Eq.\ref{unc} multiplied by $c$.
In order to obtain Hawking's radiation, 
the distance 
corresponding to 
gravitational forces promoting virtual particles to real ones cannot be 
less than $c\Delta t$. 
Assuming that $c\Delta t \sim \lambda \sim R$, 
with $\lambda$ the particle wavelength 
and $R$ the BH radius, 
we can compared it with Eq.\ref{unc}
obtaining 
\begin{equation}
	\label{compare}
	\frac{2G_{N}M_{BH}}{c^2}\sim \frac{ch}{4\pi E}\, , 
\end{equation}
which, using $E=k_{B}T_{H}$ with $k_{B}$ the Boltzmann constant, corresponds to the Hawking's temperature
\begin{equation}
	\label{coorrr}
	T_{H}\sim \frac{hc^{3}}{8\pi k_{B} G_{N} M_{BH}}\, .
\end{equation}

Now, considering superoscillating virtual pairs,
we expect an evasion of the Heisenberg's uncertainty bound 
in Eq.\ref{unc} as 
\begin{equation}
	\label{Fixxx}
	\Delta t \Delta E\sim \hbar/a\, .
\end{equation}

Repeating the previous considerations with the a-correction, 
we arrive to a modified thermal energy 
for the emission of superoscillating modes as
\begin{equation}
	\label{TTTA}
	E\sim a\frac{hc^{3}}{8\pi G_{N} M_{BH}}\sim ak_{B}T_{H}
\end{equation}
where $T_{H}$ is the standard Hawking's temperature in Eq.\ref{coorrr}.
This suggests that the BH horizon has superparticles, 
from virtual pairs in gravitational fields, 
which can escape from a BH 
with a higher energy than ordinary ones. 

The a-value can have different scales
and for small positive departure from unity (not more than a order of magnitude)
superoscillations can in principle be emitted 
in form of uniform radiation, with temperature $T\simeq a T_{H}$
with $a\sim O(1)$
and Black Body thermal distribution. 
Nevertheless, 
cases with $a>>1$ related to rare events with high energy emissions 
from a BH are not excluded but severely suppressed as $a^{-2n}$.
On the other hand, the BH entropy 
remains exactly equal to the Bekenstein-Hawking Area-law: 
the modification BH temperature of $a$-factor is compensated
by the $a$-factor multiplying the Heat $\delta Q=\delta M_{BH}$
in the formula $S_{BH}=\delta M_{BH}/T $.

Now, let us explore a semiclassical tunneling method
for superoscillating radiation, building  on the techniques suggested 
in Refs.\cite{T1,T2,T3}. 
Let us consider a Schwarzschild BH with a shell moving in the geodesic of space-time carrying (ordinary) energy $\omega$:
\begin{equation}
	\label{metr}
	ds^{2}=\Big(1-\frac{2(M-\omega)}{r} \Big)dt^{2}+2\sqrt{\frac{2(M-\omega)}{r}}dtdr+dr^{2}+r^{2}d\Omega\, ,
\end{equation}
where $M$ is the BH mass (in natural units). In Eq.\ref{metr}, we employed a set of coordinate transformations dubbed Painlev\'e ones which notably do not have any horizon singularity. 
From Eq.\ref{metr}, we can compute 
the Imaginary part of the action 
related to an s-(ordinary)wave crossing the horizon 
from a $R_{IN}$ to $R_{OUT}$:
\begin{equation}
	\label{IMS}
	{\rm Im}S=-{\rm Im}p_{r}dr= {\rm Im}\int_{0}^{\omega}\frac{1}{1-\sqrt{\frac{2(M-\tilde{\omega})}{r}}}d\tilde{\omega}\, .
\end{equation}
In case of s-superwaves,  the emission 
energy $\omega$ can be achieved from 
a lower source characteristic frequency
that is $\omega'\simeq \omega/a$ in large-n limit.

Therefore, the superoscillation case 
corresponds to 
\begin{equation}
	\label{IMS}
	{\rm Im}S= -{\rm Im}p_{r}dr\simeq {\rm Im}\int_{0}^{\omega/a}\frac{1}{1-\sqrt{\frac{2(M-\tilde{\omega})}{r}}}d\tilde{\omega}\, .
\end{equation}
with notation redefinition of $\omega'$ as $\omega$,
integrating on the frequency bandwidth delimited by the maximal Fourier frequency of the source. 

This integral can be done in complex plane from a deformed 
contour in the lower half plane of $\tilde{\omega}$.
We obtain 
\begin{equation}
	\label{ImSPRo}
	{\rm Im}S=4\pi a^{-1} \omega \Big(M-a^{-1}\frac{\omega}{2}\Big),\,\,\, R_{IN}>R_{OUT}\, . 
\end{equation}
Therefore, the transition rate in semi-classical approximation is  
\begin{equation}
	\label{Semi}
	\Gamma\simeq e^{-2{\rm Im}S}=e^{-8\pi a^{-1} \omega(M-a^{-1}\omega/2)}\, . 
\end{equation}

In case where $\omega << M$, this expression simplifies to 
\begin{equation}
	\label{Semi}
	\Gamma\sim (e^{-S_{BH}})^{1/a}\, , 
\end{equation}
where $S_{BH}$ is the Bekenstein-Hawking entropy.
Thus, the transition tunneling rate is larger than Hawking's one. 
This also implies that superoscillations 
can source deviations from BH thermality 
beyond Hawking's radiation as
\begin{equation}
	\label{Dev}
	\langle \Delta E \rangle \sim (a-1)T_{H}\,.
\end{equation}

It is worth to remark that there is an extra factor to include:
the $a^{-2n}$ suppression. 
Therefore, the transition probability is 
\begin{equation}
	\label{Semi}
	\Gamma\sim a^{-2n}e^{-2{\rm Im}S}=e^{-8\pi a^{-1} \omega M-2n\log a}\, .
\end{equation}
In the ideal case of $n\rightarrow \infty$, the transition would be totally suppressed. 
Indeed, superoscillating radiation cannot asymptotically survive for a hypothetical 
detection at the future infinite. 
However, a sufficient condition for escaping outside a BH 
is that superoscillations survive for a distance higher than 
the BH radius, corresponding to large but finite $n$. 

Let us compare the transition rate in superoscillating and ordinary settings,  \begin{equation}
	\log\left(\frac{\Gamma^{sup}}{\Gamma^{ord}}\right)\simeq \log \left( \frac{e^{-8\pi a^{-1} \omega M-2n\log a}}{e^{-8\pi  \omega M}} \right)=8 \pi  \omega M (1-a^{-1})- 2n\log a = 4 \pi \sqrt{n}   (1-a^{-1})- 2n\log a.
\end{equation}
in large n-limit approximation, 
where  we take into account the evanescence bound $\omega R_{BH} =2\omega M=\sqrt{n}$. 
\begin{figure}
	\centering
	\includegraphics[scale=1.5]{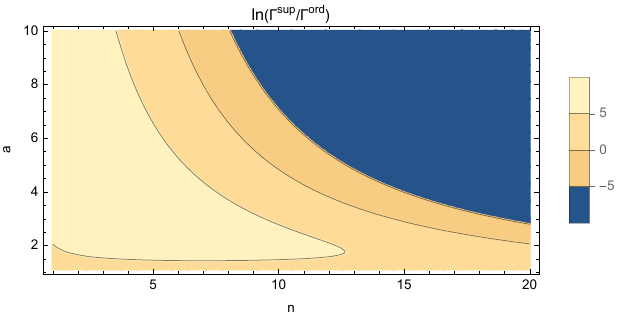}   \caption{$\log\left(\Gamma^{sup}/\Gamma^{ord}\right)$ as a function of superoscillating parameters $n$ and $a$. The positive region represents a larger  transition rate of superoscillating BH scenario compared to the ordinary one. }
	\label{figBh}
\end{figure}
As shown in the Fig. \ref{figBh}, there exists a broad parameter region where
superoscillating radiation dominates over Hawking's one, even for values of $n$ large as $n\geq 10$ (reasonably we cannot extend the result for $n\sim O(1)$ as based on a calculation in large-n approximation regime).

In principle, we can include all  
superoscillating virtual pairs 
corresponding to $\bar{a}$-parameters within the interval $[1,a]$, where we redefine $a$ as the cut-off 
excluding modes effectively too suppressed in intensity
as shown in the analysis above (see Fig. \ref{figBh}). 
Maximal deviations from Hawking's temperature predicted here are 
$\frac{\Delta T}{T_{H}} \sim a-1$.
On the other hand, as remarked above, there is no any 
corrections to the BH entropy,
in agreement with the Holographic Principle and Area-law scaling. 
Such a picture suggests a possible way out from the BH information paradox which is related to the absurdity of a perfect thermality 
and a complete lost of infalling (qu)bits. In fact, the $N$-qubits
of a in-falling generic state can be mapped into
the continuous interval of possible $a$-parameters.
Thus, superoscillating modes have potentially the power to 
carry an infinite natural number of information.
Considering a certain state with $N$-qubits, 
$|q_{1},....,q_{N}\rangle$, (for example $|0,1,1,0,...,1\rangle$)
even in the extreme limit of $N\rightarrow \infty$,
they remain a {\it countable infinite} number 
in Cantor's classification and therefore they can always be mapped to a finite set of real numbers. 
Thus, deviations from perfect thermality 
through superoscillations can carry the information 
contained in a $2^{N}$-dimensional Hilbert space. 
In particular, 
we can map the $N$-qubits $q_{1},...,q_{N}$ into superoscillations, 
within the bandwidth delimited by the Hawking's radiation frequency, with 
$\omega_{i}=a_{i}T_{H}$, $a_{i=0}=1$, 
$a_{i=N}=a$. 
We can assume equidistant frequencies for superoscillations as 
$\delta_{N}=a_{i+1}-a_{i}=(a-1)/N$.

As remarked above, such superoscillations cannot propagate for distances much larger than the BH horizon, 
thus their range is localised just near the BH radius. 
Since their origin is from quantum virtual effects, 
they provide {\it quantum hairs} on BH area. 
A countable number of quantum hairs near the BH horizon 
decorate the BH radiation in future infinity compressed in a dense compact interval 
of real numbers $[1,a]$. 
Let us estimate the information-storage capacity $I$ in form of quantum hairs from superoscillations. 
In general, the cost of a qubit storage 
in terms of energy is related to a characteristic energy gap of the system $\epsilon$ as $I\sim 1/R\epsilon$ where $R$ is the characteristic size of the system.
In case of a BH, $R$ is its radius. 
In the limit of $\epsilon<<1/R$, $I>>1$
corresponding to a large information capacitor. 
In case of superoscillations, 
$N$-qubits can be densily stored in 
a the interval $[1,a]$ 
with an inter-distance 
of $\delta_{N}=(a-1)/N$, corresponding 
to an energy discretization 
of $\Delta E_{N}\sim \delta_{N} T_{H}\sim (a-1)/(NR)$.
Thus, the energy gap to invest for one-qubit storage 
is $\epsilon \sim \Delta E_{N}$.
This implies that the information-storage capacity 
in form of superoscillations corresponds to 
$I\sim N$. Such scaling of the information capacity is  
the desired holographic one 
for BH entropy $S\sim A\sim N$, with $A$ the BH area and 
where the qubits are thought as stored
on the BH horizon.

One crucial issue of this picture is how superoscillations can transmit informations outside BHs to an asymptotic observer. 
Such a puzzle can be solved considering mixings and scatterings of 
superoscillating particles with ordinary ones.
In fact, in previous sections, we already discussed how superoscillating fields can interact with ordinary ones 
introducing interaction portal terms in the lagrangian density. 
To make a concrete example, let us consider the case of superoscillating photons 
emitted from BH horizon: i) if the BH is isolated (no external environment of particles around)
scatterings of two outcoming super-$\gamma$
can produce an electron-positron pair, or more in general 
charged fermion-antifermion pairs; or 
a superoscillating photon can transit to an ordinary one 
through a kinetic mixing 
term $-\frac{\epsilon}{4}F^{(s)}_{\mu\nu}F^{\mu\nu}$
where $\epsilon$ is the effective coupling interaction and in principle it can be generated by radiative loops;
ii) if the super-photon propagates in surrounding environment as in any realistic astrophysical situation
then it will scatter with charged particles in plasma.

Clearly, these cases are only in a subset of examples 
of the many possible interaction channels.  
As a consequence, superoscillations can transmit their informations to ordinary fields 
before vanishing; therefore superoscillations can act as {\it information mediators}
in a certain time transient. 
Let us also remark that the energy of outgoing quanta near the horizon is 
much larger than Hawking's temperature,
opening a large multiplicity of channels 
from superoscillations to ordinary particles. 
Indeed, $\Delta T(a)$ in Eq.\ref{Dev} survives for asymptotic observers in form of soft ordinary particles after s-o transferings.

Let us describe the entangled state of superoscillating fields
on BH horizon as 
\begin{equation}
	\label{superr}
	|\psi\rangle =\sum_{k'=0}^{n}\sum_{k'=0}^{n}c_{k}(n,a)c_{k'}(n,a)e^{i(1-2k/n)px}e^{i(1-2k'/n)px} \delta_{k,-k'}|k,k'\rangle\, .
\end{equation}
which can be rewritten in terms of fields acting on the Fock space.
The measurement of every k-mode collapses the measurement of specular $k'$ to a value equal to $-k$. 
One of the two tunnels outside, while the second is captured  in the BH interior;
the Complementarity Principle 
suggests that the entanglement of these pairs is preserved
even after the emission process. 
However, the evanescence bound inevitably imposes that such a superoscillating state vanishes 
outside the n-range and with it also the entanglement. 

This seems to be a possible way out to the Firewall Paradox \cite{Braunstein:2009my,Almheiri:2012rt} based on the fact 
that BH radiation cannot be entangled with both interior and exterior. 

The idea is the following.
Eq.\ref{superr} establishes a time dependent entanglement of the pair,
since such a state will disappear for a time exceeding the evanescence bound. 
The one of the pair outside transfers the information about BH interior 
to environment through interactions with SM fields before its vanishing. 
The fate of the wave function of the second of the superoscillating pair inside the BH is to vanish as well. 
Therefore, the entanglement between BH interior and environment 
decays after the transient time allowed by the evanescence bound.
Nevertheless, exterior ordinary particles carry information 
about the BH interior without any contradiction with unitarity.

Indeed the Firewall paradox is based on demonstration that the following four assumptions seem to be incompatible \cite{Braunstein:2009my,Almheiri:2012rt}:
i) the process of formation-evaporation of BH can be described by a unitary ${\bf S}$-matrix;
ii) outside BH horizon, semi-classical equations work well (including local Lorentz invariance); 
iii) the BH appears to a distant observed as a quantum system with discrete level 
with a density of state proportional to the exponential of the Bekeinstein-Hawking entropy $S_{BH}$;
iV) a free falling observer will not experience any variation (such as a firewall) when crossing BH horizon. 

As it is well know, to assume all these postulates at the same time seems to lead to 
contradictions due to the fact that early radiation is fully entangled to earlier radiation
and with BH interior at the same time. 
This situation violates several bounds such as the subaddivity of entropy: 
\begin{equation}
	\label{subb}
	S_{AB}+S_{BC}\geq S_{B}+S_{ABC}\, , 
\end{equation}
where $A$ is for early radiation mode, $B$ for later radiation, $C$ for interior modes. 

Now, $S_{BC}=0, S_{ABC}=S_{A}$ corresponds to no-firewall
and $S_{AB}<S_{A}$ indicated that the entropy of an old BH is decreasing. 
These conditions lead to the absurdum  $S_{B}\leq 0$.

However, in our case, all these Von Neumman entropies vanish after a certain transient of time,
while $S_{A},S_{B}$, related to superoscillating radiationx, are transferred to a third agent, the ordinary particles' environment $S_{D}$.
Therefore, $\Delta S_{D}=-\Delta S_{C}$
at the evanescence time.
In other words, the 
"the entanglement entropy" of BH interior and exterior is evanescent
rather than zero or "eternal". Such a metastable entanglement does not prohibit to transfer information from inside to outside the BH. 
Elaborating more on it, the $S_{AB,AC,BC}(t)=0$ for $t\geq t_{ev}$
while the principle of unitarity is preserved 
as $S_{TOT}={\rm const}$; the whole entropy is transferred to $S_{D}$ for $t> t_{ev}$. 

Accepting such a scenario implies that, the behaviour of the entanglement entropy flow,
towards information purification, is much more chaotic and always well below the Page curve \cite{Page1}. 
Indeed, in this scenario, we do not have a first stage of entanglement entropy increasing followed 
by a decreasing after crossing a critical maximal peak: our picture suggests an intermittent evolution 
of the Von Neumann entropy due to the fragility of superoscillating entanglement. 
In fact, the characteristic evanescence time $L/c\sim \omega^{-1} \sqrt{n}$
is much lower than the Page time, in turn corresponding to 
half of BH entropy evaporation. 
Thus the entanglement entropy will not grow up to one peak at the Page time
to decrease to zero later on. It will have an intermittent behaviour in time 
where one single quanta is disentangled while others just created out. Indeed, such a dynamics can lead to a non-trivial quantum chaotization around BH horizon among particle wave functions emitted in a relatively close time $t-\Delta t$, $t$ and $t+\Delta t$ and interactions with ordinary particles. 
The final entanglement entropy between the BH interior
and radiation will be zero, compatible with quantum mechanics principles.

\subsection{Information capacity and transmissions}
Superoscillations can be thought as a form of information compression, where a high-frequency signal is encoded within a lower-frequency carrier signal. This allows for the transmission of more information within a limited bandwidth.
In signal processing, superoscillations can be used for efficient data transmission and compression. By encoding information in the high-frequency components of a signal, it is possible to transmit more information using less bandwidth. In this case, the function Eq.\ref{super}
represents a superoscillating signal where the oscillations are modulated by a complex exponential function raised to the power of $a$.
This function exhibits rapid oscillations with a frequency that increases as $a$ increases. 
Thus, the superoscillating fields studied above exhibit an enhanced information transmission capacity which in principle can have 
important implications and applications. 
To determine the amount of information that can be transferred through this signal, we would consider factors such as the bandwidth of the signal and the encoding scheme used to transmit information.
The typical example proposed by {\it Berry} is how to encode a $5\times 10^3\, {\rm s}$ of $20\, {\rm KHz}$ bandwidth with Beethoven's symphony record  inside a part of $1\,{\rm Hz}$ bandlimited signal \cite{Super2}.

Let us return to the definition of the band width limited Fock space
defined from the band-limited width space, as in Eqs.\ref{FockSpa}.
Then, we can generalize the argument proposed by {\it Kempf} \cite{Kempf} for Hilbert space of band limited signals to corresponding Fock spaces.
In particular, it was shown that in a Hilbert space of band limited signals 
$H_{\omega_{max}}$, there are functions $\varphi(t)$ 
that passes through any finite number of pre-fixed points.
More specifically, given a band-width with $\omega_{max}$ 
and fixing N points $\{t_{i}\}_{i=1,..,N}$ with 
amplitude $\{A_{i}\}_{i=1,..,N}$, 
it always exists a signal that interpolates the points as
$\phi(t_{i})=A_{i}$ with $i=1,..,N$.

Here we can generalise the argument proposed in Ref.\cite{Kempf} to 
a Fock space through defining a time operator 
$\hat{t}:(f_{1},...,f_{m})=t(f_{1},...,f_{m})$ where $t$ is time eigenvalue
and extending the Shannon sample theorem,
stating that the site of time variable in a bandlimited signal is discretised 
as $t_{n+1}-t_{n}=1/2\omega_{max}$, known as the Nyquist rate. 
On the other hand, a similar argument can be used
for time varying field (Heisenberg's or alternatively interaction representation) 
and showing that it always exists a field operator 
applied to the states of bandlimited Fock space which 
can interpolate N fix points in time with arbitrary amplitudes. 

Thus in principle classical and quantum superoscillating fields such as 
gravitational or electromagnetic ones in a bandwidth range of frequencies 
$[-\omega_{max},\omega_{max}]$
can transport 
more information than ordinary ones.
However, it is worth to note that superoscillations are delicate and 
unstable against background noise. 
Indeed, from prospective of information theory, 
the Shannon-Hartley (SH) theorem provides a fundamental limit on the maximum achievable data rate for a given bandwidth and signal-to-noise ratio (SNR).
It states that the channel capacity $C$ measured in bits per second (bps), is given by
\begin{equation}
	\label{CCB}
	C=B\log_{2}(1+SNR)
\end{equation}
where $B$ is the bandwidth of the channel in hertz (Hz) and $SNR$ is the signal-to-noise ratio defined as the ratio of the signal power to the noise power.
Eq.\ref{CCB} is obtained assuming that the we have a system with white noise. 
In principle, superoscillating signals can evade the bound of Eq.\ref{CCB}. 
The cost involved entails exponentially boosting the dynamical range of the signal's amplitude relative to its amplitude resolution, in order to elevate the baud rate to bandwidth ratio, or more precisely, surpassing the noise level.
Thus, from practical purposes of information transmissions, these appear at least highly challenging,
as a subtle evasion of Nyquist bound 
\begin{equation}
	\label{NyquistR}
	\Delta t\geq \omega_{max}/4\, . 
\end{equation}

Intriguingly, considerations about information transmission 
could be related to the arguments on superoscillating radiation and BH information paradox
discussed in the previous section\footnote{It is worth to mentioned Refs.\cite{At1,At2} as previous attempts to interrelate
	BH and superoscillations, even if different than our proposals since more focus on transplanckian problems in a pre-firewall paradox prospective.}. 
Indeed superoscillating signals
could  transport a large amount of compressed information
within the BH radiation frequency bandwidth. Considering $N$ qubits, they can be carried by N superoscillations  
in a bandwidth with maximal Fourier frequency as the the
Hawking temperature. 
The information-storage capacity in superoscillations 
is $I\sim N$, corresponding to a mass gap $\epsilon\sim (a-1)(NR)^{-1}$ 
energy/qubit cost. 
For a given BH size $R$, the fastest Fourier frequency  
related to it is $\omega_{max}\sim T_{H}\sim 1/R$, 
saturating the Nyquist lower bound in Eq.\ref{NyquistR}.
Let us remark that, the considered N superoscillations, evading the Nyquist bound in BHs, 
are the ones 
below the cutoff a-parameter
corresponding to include the modes with evanescence time higher than  
$\tau \sim R/c$.

\vspace{0.2cm}

\section{Conclusions and Remarks}

In this paper, we studied the implications of
superoscillations spanning from field theories to  quantum gravity. Our objective was to lay the groundwork for a deeper understanding of superoscillations in higher energy physics. 

We began by delineating the formulation of superoscillating fields within relativistic field theories, covering scalar fields, electrodynamics, and gravity. Following this theoretical foundation, we ventured into the potential applications of superoscillations in particle scatterings and mixings, and proposed the new concept of superoscillating interferometers. These tools hold promise for advancing our understanding of the quantum gravity effects and the spacetime structure. Additionally, we explored the broader implications of superoscillations, from laboratory settings to cosmic observations, highlighting their potential in multi-messenger astronomy. Moreover, we have discussed the possible impacts of superoscillating fields in Black Hole information processing and the Firewall paradox. 

Our inquiry, however, is more suggestive of new questions than definitive answers. The implications of superoscillations in areas such as quantum gravity and the black hole information paradox remain largely uncharted territories, beckoning further explorations. Moreover, the experimental challenge lies in harnessing superoscillations effectively, akin to their application in material physics, to probe the underlying principles of Nature.

\vspace{0.5cm}

{\bf Acknowledgements}. 
AA would like to thank Massimo Bianchi, Salvatore Capozziello, Giampiero Esposito and Andrea Marini for valuable comments and suggestions. 
AA work is supported by the National Science Foundation of China (NSFC) 
through the grant No.\ 12350410358; 
the Talent Scientific Research Program of College of Physics, Sichuan University, Grant No.\ 1082204112427;
the Fostering Program in Disciplines Possessing Novel Features for Natural Science of Sichuan University, Grant No.2020SCUNL209 and 1000 Talent program of Sichuan province 2021.

\section*{Appendix A: Gupta-Bleuler quantization}

In this section, we will show how to quantize superoscillating fields in a Lorentz invariant fashion 
with the Lorentz Gauge $\partial_{\mu}A^{\mu}=0$ 
and a general lagrangian 
\begin{equation}
	\label{Lagalpha}
	\mathcal{L}=-\frac{1}{4}F_{\mu\nu}F^{\mu\nu}-\frac{1}{2\alpha}(\partial_{\mu}A^{\mu})^{2}\, .
\end{equation}
with $\alpha$ an arbitrary constant 
($\alpha=1$ corresponds to Feynman's gauge 
and $\alpha=0$ to Landau's gauge). 
The corresponding equations of motion is 
\begin{equation}
	\label{parr}
	\Box A^{\nu}=0\, . 
\end{equation}
These equations have superoscillating solutions with corresponding superoscillating quantum fields as follows:

\begin{equation}
	A^\mu(t, \boldsymbol{x})=\left.\sum_k c_k\int \frac{d^3 \boldsymbol{p}}{(2 \pi)^3} \frac{1}{2\left(1-\frac{2 k}{n}\right)^2 E} \sum_{r=0}^3\left\{b^r(\boldsymbol{p}) \epsilon_r^\mu(\boldsymbol{p}) e^{-i\left(1-\frac{2 k}{n}\right)px}+b^{r *}(\boldsymbol{p}) \epsilon_r^{* \mu}(\boldsymbol{p}) e^{i\left(1-\frac{2 k}{n}\right)px}\right\}\right|_{E=|\boldsymbol{p}|}
\end{equation}
with conjugate momenta $\pi^\mu(t, \boldsymbol{x})=\partial_t A^\mu$, 
\begin{equation}
	\pi^\mu(t, \boldsymbol{x})=i \sum_k c_k \int \frac{d^3 \boldsymbol{p}}{(2 \pi)^3} \frac{1}{2 \left(1-\frac{2 k}{n}\right)} \sum_{r=0}^3\left\{-b^r(\boldsymbol{p}) \epsilon_r^\mu(\boldsymbol{p}) e^{-i\left(1-\frac{2 k}{n}\right)px}+b^{r *}(\boldsymbol{p}) \epsilon_r^{* \mu}(\boldsymbol{p}) e^{i\left(1-\frac{2 k}{n}\right)px}\right\} \, , 
\end{equation} 

where both $\hat{\pi}^{0}=\partial \mathcal{L}/\partial \dot{\hat{A}}_{0}$ and $\hat{\pi}^{i}=\partial \mathcal{L}/\partial \dot{\hat{A}}_{i}$ are treated as dynamical. 
After the quantization, the $\partial_{\mu}\hat{A}^{\mu}=0$ constraint will reduce the degrees of freedom in the Fock space of the theory. 
Above, we define 4-polarization vectors $\epsilon_{\mu}^{\lambda}$ with $\lambda=0,1,2,3,4$ instead of 3 as in Coulomb gauge. 
The polarization 4-vectors are normalized as $\epsilon_{\mu}^{\lambda}\epsilon_{\nu}^{\lambda'}\eta_{\lambda\lambda'}=\eta_{\mu\nu}$.

We define properly normalised annihilation/creation operators as 
\begin{equation}
	\begin{aligned}
		\hat{b}^r(\mathbf{p}, k, n, a) & =\sqrt{n+1} \frac{c_k(n, a)}{\left(1-\frac{2 k}{n}\right)^2\sqrt{2 E}} \hat{b}^r(\mathbf{p}) \\
		\hat{b}^{r \dagger}(\mathbf{p}, k, n, a) & =\sqrt{n+1} \frac{c_k(n, a)}{\left(1-\frac{2 k}{n}\right)^2\sqrt{2 E}} \hat{b}^{r \dagger}(\mathbf{p}) .
	\end{aligned}
\end{equation} 
and promote the corresponding classical fields to quantum operators as
$$\hat{A}^\mu(t, \boldsymbol{x})=\sum_k \int \frac{d^3 \boldsymbol{p}}{(2 \pi)^3} \frac{1}{\sqrt{2(n+1)E}} $$
\begin{equation}
	\times \sum_{r=0}^3\left\{\hat{b}^r(\boldsymbol{p},k,n,a) \epsilon_r^\mu(\boldsymbol{p}) e^{-i\left(1-\frac{2 k}{n}\right)px}+\hat{b}^{r \dagger}(\boldsymbol{p},k,n,a) \epsilon_r^{* \mu}(\boldsymbol{p}) e^{i\left(1-\frac{2 k}{n}\right)px}\right\}
\end{equation}
$$\hat{\pi}^\mu(t, \boldsymbol{x})=i \sum_k \int \frac{d^3 \boldsymbol{p}}{(2 \pi)^3} \frac{\left(1-\frac{2 k}{n}\right)\sqrt{E}}{\sqrt{2(n+1)}}$$
\begin{equation}
	\times  \sum_{r=0}^3\left\{-\hat{b}^r(\boldsymbol{p},k,n,a) \epsilon_r^\mu(\boldsymbol{p}) e^{-i\left(1-\frac{2 k}{n}\right)px}+\hat{b}^{r \dagger}(\boldsymbol{p},k,n,a) \epsilon_r^{* \mu}(\boldsymbol{p}) e^{i\left(1-\frac{2 k}{n}\right)px}\right\} \, .
\end{equation} 

We impose the quantization conditions as 
$$[\hat{A}_{\mu}({\bf x}),\hat{A}_{\nu}({\bf y})]=[\hat{\pi}_{\mu}({\bf x}),\hat{\pi}_{\nu}({\bf y})]=0\, ,$$
\begin{equation}
	\label{quantum}
	[\hat{A}_{\mu}({\bf x}),\hat{\pi}_{\nu}({\bf y})]=i \eta_{\mu\nu}\delta^{(3)}({\bf x}-{\bf y})\,.
\end{equation}
corresponding to 
\begin{equation}
	\label{aadaggerGB}
	[\hat{b}^{\lambda}({\bf p},k,n,a),\hat{b}^{\dagger \lambda'}({\bf p'},k',n,a)]=(2\pi)^{3}\delta^{\lambda\lambda'}\delta_{k,k'}\delta^{(3)}({\bf p}-{\bf p'})\,,
\end{equation}
with $\lambda,\lambda'=1,2,3$, and 
\begin{equation}
	\label{aadirregular}
	[\hat{b}^{0}({\bf p},k,n,a),\hat{b}^{\dagger 0}({\bf p'},k',n,a)]=-(2\pi)^{3}\delta_{k,k'}\delta^{(3)}({\bf p}-{\bf p'})\, . 
\end{equation}
The Lorentz constraint 
on the Fock space $\langle \Psi|\partial^{\mu}\hat{A}_{\mu}|\Psi\rangle=0$
can be implemented through the condition
\begin{equation}
	\label{condition}
	\partial^{\mu}\hat{A}_{\mu}^{+}|\Psi\rangle=0\, ,
\end{equation}
where $\hat{A}_{\mu}=\hat{A}_{\mu}^{+}+\hat{A}_{\mu}^{-}$
and
\begin{equation}
	\hat{A}^{+\mu}(t, \boldsymbol{x})=\sum_k \int \frac{d^3 \boldsymbol{p}}{(2 \pi)^3} \frac{1}{\sqrt{2(n+1)E}} \sum_{r=0}^{3}\hat{b}^{r \dagger}(\boldsymbol{p}) \epsilon_r^{* \mu}(\boldsymbol{p}) e^{i\left(1-\frac{2 k}{n}\right)px}\, .
\end{equation}

Eq.\ref{condition} is the Gupta-Bleuler condition for superoscillating fields. 
Such a constraint demands that 
\begin{equation}
	\label{kkaakaaf}
	(\hat{b}^{3}({\bf p},k,n,a)-\hat{b}^{0}({\bf p},k,n,a))|\Psi \rangle=0\, , 
\end{equation}
where $|\Psi\rangle=|\Psi_{T}\rangle |\Phi\rangle$ 
with $|\Psi_{T}\rangle$ containing only transverse photons 
and $|\Phi\rangle$ time-like and longitudinal photons. 
Such a condition decouples the negative norm states from positive ones in the L-Fock space.


\begin{thebibliography}{10}
	
	
	\bibitem{Review}
	M. Berry {\it et al}, J. Opt. {\bf 21} (2019) 053002. 
	
	\bibitem{Super1}
	Y. Aharonov, J. Anandan, S. Popescu, L. Vaidman, Phys. Rev. Lett. {\bf 64} 2965 (1990).
	
	\bibitem{Super2}
	M. V. Berry, in Proc. {\it Intl. Conf. on Fund. Aspects of Quantum Theory, Columbia}, SC, USA, 10-12 Dec. 1992 Eds. J.S. Anandan, J. L. Safko, World Scientific, Singapore (1995).
	
	\bibitem{Super3}
	M. V. Berry, J. Phys. A {\bf 27} L391 (1994)
	
	\bibitem{Super4}
	Y. Aharonov, B. Reznik, A. Stern, Phys. Rev. Lett. {\bf 81} 2190 (1998).
	
	
	\bibitem{Osc1}
	R. Buniy, F. Colombo, I. Sabadini and D.C. Struppa, 
	J. Math. Phys. {\bf 55} 113511 (2014) .
	
	
	\bibitem{Osc2}
	F. Colombo, J. Gantner and D.C. Struppa,  
	J. Math. Phys. {\bf 58} 092013 (2017).
	
	\bibitem{BesselJJJ}
	M.R. Dennis, A.C. Hamilton, J. Courtial, 
	Opt. Lett. 33 2976–8 (2018).
	
	\bibitem{coss}
	R. Remez, A. Arie,
	Optica 2 472–5 (2015).
	
	\bibitem{Antenna}
	A.M.H. Wong, G.V. Eleftheriades,
	IEEE Trans. Antennas Propag. 59 4766–76 (2011). 
	
	\bibitem{APR}
	Aharonov Y, Popescu S and Rohrlich D {\bf 1991} 
	{\it How can an infrared photon behave as a gamma ray?} Tel-Aviv University Preprint TAUP 1847-90.
	
	\bibitem{APR2}
	Aharonov, Y., Popescu, S., Rohrlich, D.
	{\it On conservation laws in quantum mechanics}, arXiv e-printsarXiv: 1609.05041 [quant-ph] (2016). 
	
	\bibitem{Fock1}
	V. Fock, Z. Phys. , 75 (1932) pp. 622–647.
	
	\bibitem{Fock2}
	F.A. Berezin, 
	Acad. Press (1966) (Translated from Russian) (Revised (augmented) second edition: Kluwer, 1989).
	
	\bibitem{GS1} 
	D. M. Eardley and S. B. Giddings, 
	Phys. Rev. D {\bf 66}, 044011 (2002).
	
	\bibitem{GS2}
	E. Kohlprath and G. Veneziano, 
	JHEP {\bf 0206} (2002) 057 [gr-qc/0203093].
	
	\bibitem{AC}
	G. Amelino-Camelia, Nature {\bf 398}, 216 (1999).
	
	\bibitem{aLIGO1}
	J Aasi, B P Abbott, R Abbott, T Abbott, M R Abernathy, K Ackley, C Adams, T Adams, P Addesso, R X Adhikari, {\it et al.}, 
	Classical and Quantum Gravity, {\bf 32}(7):074001, mar 2015.
	
	\bibitem{aLIGO2}
	B.P. Abbott {\it et al.}, 
	Phys. Rev. Lett., {\bf 116}(13):131103, 2016.
	
	\bibitem{LISA}
	Amaro-Seoane, Pau, {\it et al.}, 
	"Laser interferometer space antenna."
	arXiv preprint arXiv:1702.00786 (2017).
	
	
	\bibitem{MHZ1}
	A. M. Cruise. 
	Class. Quant. Grav., 29:095003, 2012.
	
	
	\bibitem{Aggarwal:2020olq}
	N.~Aggarwal, O.~D.~Aguiar, A.~Bauswein, G.~Cella, S.~Clesse, A.~M.~Cruise, V.~Domcke, D.~G.~Figueroa, A.~Geraci and M.~Goryachev, \textit{et al.}
	Living Rev. Rel. \textbf{24} (2021) no.1, 4
	doi:10.1007/s41114-021-00032-5
	[arXiv:2011.12414 [gr-qc]].
	
	\bibitem{Bianchi:2008pu}
	M.~Bianchi, H.~Elvang and D.~Z.~Freedman,
	JHEP \textbf{09} (2008), 063
	doi:10.1088/1126-6708/2008/09/063
	[arXiv:0805.0757 [hep-th]].
	
	\bibitem{Marletto:2017kzi}
	C.~Marletto and V.~Vedral,
	Phys. Rev. Lett. \textbf{119} (2017) no.24, 240402
	doi:10.1103/PhysRevLett.119.240402
	[arXiv:1707.06036 [quant-ph]].
	
	\bibitem{Bose:2017nin}
	S.~Bose, A.~Mazumdar, G.~W.~Morley, H.~Ulbricht, M.~Toro\v{s}, M.~Paternostro, A.~Geraci, P.~Barker, M.~S.~Kim and G.~Milburn,
	Phys. Rev. Lett. \textbf{119} (2017) no.24, 240401
	doi:10.1103/PhysRevLett.119.240401
	[arXiv:1707.06050 [quant-ph]].
	
	\bibitem{vandeKamp:2020rqh}
	T.~W.~van de Kamp, R.~J.~Marshman, S.~Bose and A.~Mazumdar,
	Phys. Rev. A \textbf{102} (2020) no.6, 062807
	doi:10.1103/PhysRevA.102.062807
	[arXiv:2006.06931 [quant-ph]].
	
	\bibitem{LIGOsq}
	M. Tse {\it et al.} (LIGO collaboration), 
	Phys. Rev. Lett. {\bf 123}, 231107, 
	https://doi.org/10.1103/PhysRevLett.123.231107. 
	
	
	
	\bibitem{Walls}
	D. Walls, 
	Nature {\bf 306}, 141-146 (1983), 
	https://doi.org/10.1038/306141a0.
	
	\bibitem{ET1}
	M. Punturo {\it et al.}, 
	Quant. Grav. {\bf 27} (2010) 194002.
	
	\bibitem{ET2}
	S. Hild {\it et al.}, 
	Class. Quant. Grav. {\bf 28} (2011) 094013, arXiv:1012.0908 [gr-qc].
	
	\bibitem{CE1}
	D. Reitze {\it et al.}, 
	Bull. Am. Astron. Soc. 51 no. 7, (2019) 035, arXiv:1907.04833 [astro-ph.IM].
	
	\bibitem{CE2}
	M. Evans et al., “A Horizon Study for Cosmic Explorer: Science, Observatories, and Community,” arXiv:2109.09882 [astro-ph.IM].
	
	\bibitem{Mpaper} 
	V. F. Mukhanov,  "Quantum Theory of Gauge Invariant Cosmological Perturbations," Soviet Physics JETP, vol. 67, no. 7, 1985, pp. 1297-1302. 
	
	\bibitem{Spaper}
	M. Sasaki, "Gauge Invariant Scalar Perturbations in the New Inflationary Universe," Progress of Theoretical Physics, vol. 70, no. 2, 1983, pp. 394-411.
	
	\bibitem{PDGNeutrino}
	M.C. Gonzalez-Garcia, M. Yokoyama, 14. Neutrino Masses, Mixings and Oscillations,
	Particle Data Group,
	https://pdg.lbl.gov/2020/reviews/rpp2020-rev-neutrino-mixing.pdf.
	
	\bibitem{EllisCPT}
	J.R. Ellis, J.S. Hagelin, D.V. Nanopoulos and M. Srednicki, Search for violations of quantum mechanics, Nucl. Phys. B {\bf 241} (1984) 381.
	
	\bibitem{Amelino-Camelia:2000ikd}
	G.~Amelino-Camelia and F.~Buccella,
	Mod. Phys. Lett. A \textbf{15} (2000), 2119-2128
	doi:10.1142/S0217732300002474
	[arXiv:hep-ph/0001305 [hep-ph]].
	
	\bibitem{Addazi:2017bbg}
	A.~Addazi, P.~Belli, R.~Bernabei and A.~Marciano,
	Chin. Phys. C \textbf{42} (2018) no.9, 094001
	doi:10.1088/1674-1137/42/9/094001
	[arXiv:1712.08082 [hep-th]].
	
	\bibitem{Addazi:2021xuf}
	A.~Addazi, J.~Alvarez-Muniz, R.~Alves Batista, G.~Amelino-Camelia, V.~Antonelli, M.~Arzano, M.~Asorey, J.~L.~Atteia, S.~Bahamonde and F.~Bajardi, \textit{et al.}
	Prog. Part. Nucl. Phys. \textbf{125} (2022), 103948
	doi:10.1016/j.ppnp.2022.103948
	[arXiv:2111.05659 [hep-ph]].
	
	
	\bibitem{DAFNE}
	L. Maiani, G. Pancheri, N. Paver, eds., \textit{The Second DA\(\Phi\)NE Physics Handbook}, INFN, Laboratori Nazionali di Frascati, 1995.
	
	\bibitem{DiDomenico}
	A. Di Domenico, ``Kaon Interferometry: a Quantum Test of the Equivalence Principle,'' \textit{Foundations of Physics}, Vol. 40, Issue 8, pp. 852-866 (2010).
	
	\bibitem{CPLEAR}
	A. Angelopoulos et al, $K_{0}-\bar{K}_{0}$ mass and decay-width differences: CPLEAR
	evaluation, report CERN-EP/99-150, Phys. Lett. B 471 (1999) 332-8.
	
	\bibitem{NA}
	R. Carosi et al, Phys. Lett. B{\bf 237} (1990) 303.
	
	\bibitem{E7}
	M. Woods et al, Phys. Rev. Lett. {\bf 60} (1988) 1695; J.R. Patterson et al, Phys. Rev. Lett. {\bf 64} (1990) 1491.
	
	
	\bibitem{KLOE}
	KLOE-2 Collaboration, ``Precision tests of discrete symmetries in the \(K^0\) system with the KLOE detector,'' \textit{The European Physical Journal C} (2014) 74: 3198.
	
	\bibitem{Addazi:2016yre}
	A.~Addazi,
	Int. J. Geom. Meth. Mod. Phys. \textbf{14} (2016) no.01, 1750012
	doi:10.1142/S0219887817500128
	[arXiv:1607.02593 [hep-th]].
	
	
	\bibitem{Superluminal1}
	A.M. Steinberg, 1995, Phys. Rev. A 52, 32.
	
	\bibitem{Superluminal2}
	A.M. Steinberg, 1995, Phys. Rev. Lett. 74, 2405.
	
	\bibitem{Superluminal3}
	V. S.Olkhovsky, E. Recami, G. Salesi, 
	Eur. Phys. Letters, {\bf 57} (2002) 879-884,
	https://doi.org/10.48550/arXiv.quant-ph/0002022, 
	arXiv:quant-ph/0002022.
	
	\bibitem{Superluminal4}
	Y. Aharonov, N. Erez, B. Reznik, Phys. Rev. A 65, 052124 (2002), 
	https://doi.org/10.48550/arXiv.quant-ph/0110104
	[arXiv:quant-ph/0110104] 
	
	\bibitem{Hawking1}
	S.W. Hawking, Commun. Math. Phys. {\bf 43} (1975) 199.
	
	\bibitem{Hawking2}
	S. W. Hawking, 
	Phys. Rev. D{\bf 14}, 2460-2473 (1976)
	
	\bibitem{Pino}
	J. Pinochet, 
	Phys. Educ. {\bf  51} (2016) 015010.
	
	\bibitem{T1}
	E. Keski-Vakkuri and P. Kraus, Phys. Rev. D{\bf 54} (1996) 7407.
	
	\bibitem{T2}
	P. Kraus and F. Wilczek, Nucl. Phys. B{\bf 433} (1995) 403.
	
	\bibitem{T3}
	M.K. Parikh and F. Wilczek, Phys. Rev. Lett. {\bf 85} (2000) 5042.
	
	
	\bibitem{Susskind:1993if}
	L.~Susskind, L.~Thorlacius and J.~Uglum,
	Phys. Rev. D \textbf{48} (1993), 3743-3761
	doi:10.1103/PhysRevD.48.3743
	[arXiv:hep-th/9306069 [hep-th]].
	
	\bibitem{Braunstein:2009my}
	S.~L.~Braunstein, S.~Pirandola and K.~\.Zyczkowski,
	Phys. Rev. Lett. \textbf{110} (2013) no.10, 101301
	doi:10.1103/PhysRevLett.110.101301
	[arXiv:0907.1190 [quant-ph]].
	
	\bibitem{Almheiri:2012rt}
	A.~Almheiri, D.~Marolf, J.~Polchinski and J.~Sully,
	JHEP \textbf{02} (2013), 062
	doi:10.1007/JHEP02(2013)062
	[arXiv:1207.3123 [hep-th]].
	
	\bibitem{Page1}
	Don N. Page, 
	Phys. Rev. Lett. {\bf 71}, 3743–3746 (1993),
	arXiv:hep-th/9306083 [hep-th]
	
	\bibitem{Kempf}
	A. Kempf 
	J. Math. Phys. {\bf 41} 2360 (2000). 
	
	\bibitem{At1}
	H. Rosu, Nuovo Cim. 112B 131 (1997), gr-qc/9606070.
	
	\bibitem{At2}
	B. Reznik, Phys. Rev. D{\bf 55} 2152 (1997)
	
	
	
	
\end{thebibliography}
\end{document}